\documentclass[twocol]{ametsoc}
\usepackage{ulem}

\journal{jpo}

\bibpunct{(}{)}{;}{a}{}{,}

\title{On Galerkin approximations of the surface-active quasigeostrophic equations}

\authors{Cesar B. Rocha\correspondingauthor{Cesar B. Rocha, Scripps Institution of Oceanography, University of California, San Diego, 9500 Gilman Dr. MC 0213, La Jolla, CA 92093} and William R. Young}
\affiliation{Scripps Institution of Oceanography, University of California, San Diego, La Jolla, California}
\extraauthor{Ian Grooms}
\extraaffil{Center for Atmosphere Ocean Science, Courant Institute of Mathematical Sciences, New York University, New York, New York}

\email{crocha@ucsd.edu}

\newcommand{\com}{\, ,}
\newcommand{\per}{\, .}

\newcommand{\defn}{\ensuremath{\stackrel{\mathrm{def}}{=}}}

\def\beq{\begin{equation}}
\def\eeq{\end{equation}}

\newcommand{\p}{\partial}
\newcommand{\ii}{{\rm i}}
\newcommand{\dd}{{\rm d}}

\newcommand{\al}{\alpha}

\newcommand{\half}{\tfrac{1}{2}}
\newcommand{\halfrho}{\tfrac{1}{2}}
\newcommand{\rz}{{}}

\newcommand{\helmn}{\triangle_n}
\newcommand{\helms}{\triangle_s}

\newcommand{\sA}{\mathsf{A}}
\newcommand{\sB}{\mathsf{B}}

\newcommand{\sJ}{\mathsf{J}}

\newcommand{\sP}{\mathsf{P}}

\newcommand{\sL}{\mathsf{L}}
\renewcommand{\sJ}{\mathsf{J}}

\newcommand{\sT}{\mathsf{T}}

\newcommand{\sC}{\mathsf{C}}

\newcommand{\bcdot}{\hspace{-0.1em} \boldsymbol{\cdot} \hspace{-0.12em}}
\newcommand{\bnabla}{\boldsymbol{\nabla}}

\newcommand{\grad}{\bnabla}

\newcommand{\lap}{\triangle}

\newcommand{\vth}{\vartheta}

\newcommand{\vthb}{\vartheta^{\mathrm{-}}}
\newcommand{\vthbhat}{{\hat{\vartheta}}^{\mathrm{-}}}

\newcommand{\psib}{\psi^{\mathrm{-}}}

\newcommand{\vtht}{\vartheta^{\mathrm{+}}}
\newcommand{\vththat}{{\hat{\vartheta}}^{\mathrm{+}}}
\newcommand{\vthtbhat}{{\hat{\vartheta}}^{\pm}}

\newcommand{\vthtb}{\vartheta^{\pm}}
\newcommand{\vThtb}{\varTheta^{\pm}}

\newcommand{\gpsi}{\breve \psi}
\newcommand{\gphi}{\breve \phi}
\newcommand{\gq}{\breve q}
\newcommand{\gU}{\breve U}
\newcommand{\gQ}{\breve Q}
\newcommand{\gsigma}{\breve \sigma}

\newcommand{\psit}{\psi^{\mathrm{+}}}

\newcommand{\bur}{\left(\tfrac{f_0}{N}\right)^2}

\renewcommand{\sp}{\mathsf{p}}

\newcommand{\nmax}{\mathrm{N}}

\newcommand{\sq}{\mathsf{q}}
\newcommand{\cosech}{\text{csch}\,}

\newcommand{\zp}{z^+}
\newcommand{\zm}{z^-}
\newcommand{\qA}{q^A_{\nmax}}
\newcommand{\psiB}{\psi^B_{\nmax}}
\newcommand{\phiB}{\phi^B_{\nmax}}

\newcommand{\psiG}{\psi^{\mathrm{G}}}
\newcommand{\qG}{q^{\mathrm{G}}}

\newcommand{\UG}{U^{\mathrm{G}}}
\newcommand{\UGN}{U^{\mathrm{G}}_{\nmax}}

\usepackage{mathtools}
\usepackage{graphicx}

\abstract{We study the representation of  solutions of the three-dimensional quasigeostrophic (QG) equations using Galerkin series with standard vertical modes, with particular attention to the incorporation of active surface buoyancy dynamics.  We extend two existing Galerkin approaches (A and B) and develop a new Galerkin approximation (C). Approximation A, due to \cite{flierl1978}, represents the streamfunction as a truncated Galerkin series and defines the potential vorticity (PV) that satisfies the inversion problem exactly. Approximation B, due to \cite{tulloch_smith2009b}, represents the PV as a truncated Galerkin series and calculates the streamfunction that satisfies the inversion problem exactly. Approximation C, the true Galerkin approximation for the QG equations, represents both streamfunction and PV as truncated Galerkin series, but does not satisfy the inversion equation exactly. The three approximations are fundamentally different unless the boundaries are isopycnal surfaces.  We discuss the advantages and limitations of approximations A, B, and C in terms of mathematical rigor and conservation laws, and illustrate their relative efficiency by solving linear stability problems with nonzero surface buoyancy.  With moderate number of modes, B and C have have superior accuracy than A at high wavenumbers. Because B lacks conservation of energy, we recommend approximation C for constructing   solutions to the surface-active QG equations using Galerkin series with standard vertical modes.}

\begin{document}

\maketitle

%


\section{Introduction}
\label{intro_sec}

Recent interest in upper-ocean dynamics and sub-mesoscale turbulence  has focussed attention on surface geostrophic dynamics and the role of surface   buoyancy variations.  A main issue is the representation of  active surface buoyancy by finite vertical truncations of the quasigeostrophic (QG) equations. 
 Standard multi-layer \citep[e.g., ][]{pedlosky1987} and  modal approximations \citep[e.g., ][]{flierl1978}
 assume  that there is no variation of buoyancy on the surfaces.

 Here we explore the representation of surface and interior  dynamics using the simple and familiar  vertical modes of physical oceanography. These modes, denoted here by $\sp_n(z)$, are defined by the  Sturm-Liouville eigenproblem
\beq
\frac{\dd }{\dd z} \frac{f_0^2}{N^2} \frac{\dd  \sp_n }{\dd z} = - \kappa_n^2\sp_n\com
\label{SL1}
\eeq
with homogeneous Neumann boundary conditions at the bottom ($z=z^-$) and top ($z=z^+$) surfaces of the domain: 
\beq
 \frac{\dd \sp_n}{\dd z}\left(z^\pm\right)=0 \per
 \label{SL2}
 \eeq
  In \eqref{SL1} $N$ is the buoyancy frequency and $f_0$ is the Coriolis parameter.  The eigenvalue $\kappa_n$ in \eqref{SL1} is the deformation wavenumber of the $n$'th mode. With normalization, the modes satisfy the orthogonality condition
\beq
\tfrac{1}{h} \int_{z^-}^{z^+} \!\!\! \sp_n\sp_m \, \dd z = \delta_{mn} \com
\eeq
where $h\defn \zp-\zm$ is the depth. The  barotropic mode  is $\sp_0=1$ and $\kappa_0=0$.


The  modes defined by the eigenproblem \eqref{SL1} and \eqref{SL2} provide a fundamental basis for representing  solutions of both the primitive and quasigeostrophic equations  as a linear combination of  $\lbrace \sp_n \rbrace$ \citep{gill1982,pedlosky1987,vallis2006,ferrari_wunsch2010,lacasce2012}. In fact,  the set $\lbrace \sp_n \rbrace$ is mathematically complete and can be used to represent \textit{any} field with finite square integral,
\beq
\int_{z^-}^{z^+} \!\!\!\! \phi^2 \, \dd z < \infty\per
\label{finiteL2}
\eeq
Even if the field $\phi$ has nonzero derivative at $z^{\pm}$, or internal discontinuities,  its representation as a linear combination of  the basis functions $\lbrace \sp_n \rbrace$ converges in $L^2(z^-,z^+)$ i.e., the integral of the squared error goes to zero as the number of basis functions increases \citep[e.g.,][ch.~10]{HN2001}. 

Despite the rigorous assurance of completeness in the previous paragraph,  the utility of $\lbrace \sp_n \rbrace$  for problems with nonuniform surface buoyancy has been questioned by several authors  \citep[e.g., ][]{lapeyre2009,roullet_etal2012,smith_vanneste2013}. These authors   argue that  the  homogeneous boundary conditions  in  \eqref{SL2} are incompatible with non-zero surface buoyancy and   that  representation of the streamfunction $\psi$ as a linear combination of $\lbrace \sp_n \rbrace$  is useless if $\psi_z$ is non-zero on the surfaces. This supposed  incompatibility of \eqref{SL2} with non-zero surface buoyancy  is a  main motivation for a new, alternative set of orthogonal basis functions proposed by   \cite{smith_vanneste2013}.


The aim of this paper is to obtain a good Galerkin approximation to solutions of the QG equation with non-zero surface buoyancy  using the familiar basis $\lbrace \sp_n \rbrace$. We show that that both the inversion problem \textit{and} evolutionary dynamics can be handled using $\lbrace \sp_n \rbrace$ to represent the streamfunction. As part of this program we revisit and extend two existing modal approximations \citep{flierl1978,tulloch_smith2009a}, and develop a new Galerkin approximation. We discuss the relative merit of the three approximations in terms of their mathematical rigor and conservation laws, and illustrate their efficiency and caveats by solving linear stability problems with nonzero surface buoyancy.

Using concrete examples, we show that the concerns expressed by earlier authors regarding the suitability of the standard modes $\lbrace \sp_n \rbrace$ are over-stated: even with non-zero surface buoyancy, the Galerkin expansion  for $\psi$ in terms of $\lbrace \sp_n \rbrace$ converges absolutely and uniformly with no Gibbs phenomena.  A modest number of terms provides a good approximation to $\psi$ throughout the domain, including on  the top and bottom boundaries. In other words,   the surface streamfunction can be expanded in terms of  $\lbrace \sp_n \rbrace$ and, with enough modes,  this representation can then be used to accurately calculate  the  advection of non-zero surface buoyancy. In section \ref{eadysec} we illustrate this procedure by solving the classic Eady problem using the basis $\lbrace \sp_n \rbrace$ for the streamfunction.



\section{The exact system}
\label{exact_qg_sec}

In this section we summarize the basic properties of the QG system. For a detailed derivation see \cite{pedlosky1987}.

\subsection{Formulation}
The streamfunction is denoted $\psi(x,y,z,t)$ and we  use the following notation. 
 \beq
 u = -\psi_y\com \qquad v = \psi_x\com \qquad \vth = \bur \psi_z \per
 \eeq
 The variable  $\vth$ is related to the  buoyancy by  $b = N^2\vth/f_0$. The QG potential vorticity (QGPV) equation is
\beq
\p_t q + \sJ (\psi,q) + \beta v=0\com
\label{QGPV1}
\eeq
where the potential vorticity is
 \beq
 q = \left(\lap + \sL\right) \psi \com
 \eeq
 with 
 \beq
 \lap \defn \p_x^2+\p_y^2\com \qquad \text{and} \qquad \sL \defn \p_z \bur\p_z \per
 \eeq 
Also in \eqref{QGPV1}, the Jacobian is $\sJ(A,B) \defn \p_x A \, \p_y B - \p_y A \, \p_x B$.

 The boundary conditions at the top ($z=\zp$) and  bottom ($z=\zm$) are that $w=0$, or equivalently
 \beq
 @z=z^{\pm}\,:\qquad \p_t \vth^{\pm} + \sJ(\psi^{\pm},\vth^{\pm})=0\per \label{bc1}
 \eeq
 Above we have used the superscripts $+$ and $-$ to denote evaluation at $\zp$ and $\zm$ e.g., $\psi^+ = \psi(x,y,\zp,t)$.

\subsection{Quadratic conservation laws}
 In the absence of sources and sinks, the exact QG system has four quadratic conservation laws:   energy, potential enstrophy, and surface buoyancy variance at the two surfaces \citep[e.g., ][]{pedlosky1987,vallis2006}. Throughout we assume horizontal periodic boundary conditions.

The well-known energy conservation law is
\beq
\label{cons_E_exact}
\frac{d E}{dt} = 0 \com 
\eeq
where 
\beq
 \label{E_defn}
 E\defn \int \halfrho |\grad \psi|^2 +  \halfrho  \bur  \!\left(\p_z \psi\right)^2\, \dd V \per
 \eeq
The total energy is $\rho_0\,E$, where $\rho_0$ is a reference density. An alternative expression for $E$ is
 \beq
 E =- \halfrho \int \!\! \psi q\, \dd V + \halfrho \int \! \psit \vtht- \psib \vthb\dd S \per
  \label{PE7}
\eeq
If $q=0$ (e.g., as in the Eady  problem) then \eqref{PE7}  expresses $E$ in terms of surface contributions.

If $\beta =0$ then there are many quadratic potential enstrophy invariants: the volume integral of $q^2 A(z)$, with   $A(z)$  an arbitrary function of the vertical coordinate, is conserved. The choice $A(z) = \delta(z-z_*)$ reduces to conservation of the surface integral of $q^2$ at any level $z_*$. 

\cite{charney1971} noted that, in a doubly periodic domain,  nonzero $\beta$ destroys all these quadratic potential enstrophy conservation laws, including the conservation of potential enstrophy defined simply as the volume integral of $q^2$. Multiplying the QGPV equation \eqref{QGPV1} by $q$, and integrating by parts, we obtain 
\beq
\frac{\dd}{\dd t} \int \half q^2 \, \dd V + \beta  \int \big[v \vartheta \big]_{z^-}^{z^+} \, \dd S =0\per
\label{EP7}
\eeq
The potential enstrophy equation \eqref{EP7} is the finite-depth analog of Charney's equation (13). To make progress Charney assumed $\vth = 0$ at the ground. But the $\beta$-term on the right of \eqref{EP7} can be eliminated by cross-multiplying the QGPV equation  \eqref{QGPV1} evaluated at the surfaces $z^{\pm}$ with the boundary conditions \eqref{bc1}, and combining with \eqref{EP7}. Thus nonzero $\beta$ selects a uniquely conserved potential enstrophy from the infinitude of $\beta=0$ potential enstrophy conservation laws:
\beq
\frac{d Z}{dt} = 0 \com 
\label{cons_Z_exact}
\eeq
where the potential enstrophy is
\beq
Z\defn \int \half q^2 \, \dd V - \int q^+\vtht - q^- \vthb \, \dd S \per
\label{ensdens}
\eeq
With $\beta \neq 0$ the surface contributions in \eqref{ensdens} are required to form a conserved quadratic quantity involving $q^2$. Notice that \eqref{ensdens} is not sign-definite. To our knowledge, the  conservation law in \eqref{cons_Z_exact} and   \eqref{ensdens} is previously unremarked.
 
Finally, in addition to $E$ and $Z$, the surface buoyancy variance is conserved on each surface
\beq
\label{sB_cons}
\frac{\dd}{\dd t} \int \half \left( \vartheta^{\pm}\right)^2 \dd S=0\per
\eeq
Thus, with $\beta \neq 0$, the QG model has four quadratic conservation laws: $E$, $Z$ and the buoyancy variance at the two surfaces. 

\section{Galerkin approximation using standard vertical modes}
\label{galerk_sec}  
A straightforward approach is to represent the streamfunction  by linearly combining the first  $\nmax+1$ vertical modes.  The mean square error in this approximation  is
\beq
\text{err}_{\psi}(a_0, a_1, \cdots a_\nmax)  \defn \frac {1}{h} \int_{z^-}^{z^+} \Big( \psi - \sum_{n=0}^{\nmax} a_n \sp_n\Big)^2 \, \dd z \per
\label{msqe}
\eeq
We use a roman font, and context, to distinguish the truncation index $\nmax$ in \eqref{msqe} from the buoyancy frequency $N(z)$.
The coefficients $a_0$ through $a_\nmax$ are determined to minimize  $\text{err}_{\psi}$, and thus one obtains  the Galerkin approximation $\psiG_{\nmax}$  to the exact streamfunction:
\beq
\psiG_{\nmax}(x,y,z,t)\defn \sum_{n=0}^{\nmax} \gpsi_n(x,y,t) \sp_n(z)\com
\label{straight8}
\eeq
where the coefficients in the sum above are
\beq
\gpsi_n(x,y,t) \defn \tfrac{1}{h} \int_{z^-}^{z^+} \!\!\!\! \psi\, \sp_n \, \dd z\, .
\label{straight9}
\eeq
Throughout we use the superscript $\breve{}$ to denote a Galerkin coefficient defined via projection of a field  onto a vertical mode.

In complete analogy with the streamfunction, one can also develop an $(\nmax+1)$-mode Galerkin approximation to   the PV:
\beq
\qG_{\nmax}(x,y,z,t)\defn  \sum_{n=0}^{\nmax} \gq_n(x,y,t) \sp_n(z)\com
\label{straight13}
\eeq
with coefficients
\beq
\gq_n \defn \tfrac{1}{h} \int_{z^-}^{z^+} \!\!\!\! q\, \sp_n \, \dd z \per
\label{straight17}
\eeq
The construction of the Galerkin approximation $\qG_{\nmax}$ above minimizes a mean square error $\text{err}_q$ defined in analogy with \eqref{msqe}.

Now recall that  the exact $\psi$ and $q$ are related by the  elliptic ``inversion problem":
\beq
(\lap + \sL) \psi = q\com 
\label{inv3}
\eeq
with boundary conditions  at $z^{\pm}$:
\beq
  \bur \psi_z = \vartheta^{\pm}\per
  \label{inv5}
\eeq
The Galerkin approximations in \eqref{straight8} through \eqref{straight17} are defined independently of the information in \eqref{inv3} and \eqref{inv5}.
The relationship between the Galerkin coefficients $\gq_n$ and $\gpsi_n$ is obtained by
multiplying \eqref{inv3} by $\tfrac{1}{h}\sp_n(z)$ and integrating over the depth. Noting the intermediate result
\beq
\tfrac{1}{h}\int_{\zm}^{\zp}\!\!\! \sp_n \sL \psi \, \dd z = \tfrac{1}{h}\left[ \sp_n^+ \vtht  - \sp_n^- \vthb\right] - \kappa_n^2  \gpsi_n\com
\eeq
 we obtain
 \beq
\gq_n =  \helmn \gpsi_n+\underbrace{ \tfrac{1}{h}   \left( \sp_n^+\,  \vtht  -  \sp_n^-\, \vthb \right)}_{\text{surface terms}} \com
 \label{sol11}
 \eeq
 where $\helmn$ is the $n$'th mode  Helmholtz operator
 \beq
 \helmn \defn \lap - \kappa^2_n\per
 \label{helmdef}
 \eeq
The relation in \eqref{sol11} is the key to  a good Galerkin approximation to surface-active quasigeostrophic dyamics.

 Term-by-term differentiation of the $\psiG_{\nmax}$-series in \eqref{straight8} does not give the $\qG_{\nmax}$ series in \eqref{straight13} unless $\vartheta^\pm=0$. In other words, term-by-term differentiation does not produce the correct  relation \eqref{sol11} between  $\gq_n$ and $\gpsi_n$.  Thus the Galerkin truncated PV and the Galerkin truncated streamfunction do not satisfy the inversion boundary value problem  exactly
 \beq
 \big( \lap +\sL\big) \psiG_{\nmax} \neq \qG_{\nmax}\per
 \label{unpleasant}
 \eeq	
Despite \eqref{unpleasant}, the truncated series $\psiG_{\nmax}$ and $\qG_{\nmax}$ are the best least-squares approximations to $\psi$ and $q$.

Notice that, in analogy with the Galerkin approximations for $q$ and $\psi$,
\beq
\label{delta_brev}
\breve \delta^{+}_n = \tfrac{1}{h}\sp_n^{+} \qquad \text{and} \qquad \breve \delta^{-}_n = \tfrac{1}{h}\sp_n^{-}\com
\eeq
where 
\beq
\label{series_delta_brev}
\delta_{\nmax}^{+G}(z) = \sum_{n=0}^{\nmax} \breve \delta^{+}_n \sp_n \qquad \text{and} \qquad \delta_{\nmax}^{-G}(z) = \sum_{n=0}^{\nmax} \breve \delta^{-}_n\sp_n \com
\eeq
are finite approximations to distributions $\delta(z-z^{\pm})$ at the surfaces. Of course, these surface $\delta$-distributions do not satisfy the $L^2$ convergence condition in \eqref{finiteL2} and thus  the series in \eqref{series_delta_brev} only converge in a distributional sense.\citep[e.g.,][]{HN2001}. For instance, if $\phi$ satisfies the $L^2$ convergence condition in \eqref{finiteL2}, then
\beq
\int_{\zm}^{\zp}\!\! \phi(z) \delta_{\nmax}^{+G}(z)\, \dd z\,\, \to \,\int_{\zm}^{\zp}\!\! \phi(z) \delta(z-\zp)\, \dd z = \phi(\zp)\com
\eeq
as $\nmax \to \infty$. Thus, in that limit, 
\beq
 \big( \lap +\sL\big) \psiG_{\nmax} \rightharpoonup q-  \delta(z-z^+ )\,\vtht +  \delta(z-z^- )\,\vthb \com
 \label{Bretherton0}
 \eeq	
where $\rightharpoonup$ denotes distributional convergence. The right-hand-side of \eqref{Bretherton0} is the Brethertonian modified potential vorticity  \citep{bretherton1966} with the boundary conditions incorporated as PV sheets. To illustrate \eqref{unpleasant} and \eqref{Bretherton0} we present an elementary example that is is relevant to our discussion of the Eady problem in section \ref{eadysec}. 
 
\subsection*{An elementary example: the Eady basic state}

\begin{figure*}
\begin{center}
\includegraphics[width=38pc,angle=0]{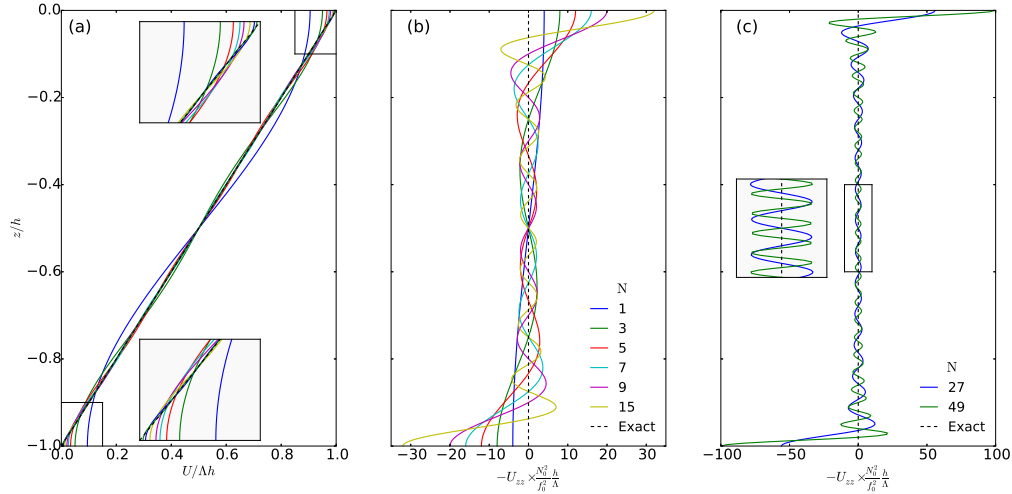} \\
\end{center}
\caption{Nondimensional base-state for the Eady problem using various truncation for the series \eqref{easyexact2}. In the middle panel $\nmax$ is the number of baroclinic modes. (a) Zonal velocity: although   the truncation has zero slope at the boundaries there are no Gibbs oscillations. (b) Meridional PV  gradient  associated with the truncated series \eqref{easyexact4}. (c) as in (b) but  with an expanded abscissa. As $\nmax$ increases, the PV gradient distributionally converges to two Brethertonian delta functions at the boundaries.}
\label{eady_mean_state}
\end{figure*}

As an example, consider the case with constant buoyancy frequency $N$. We use nondimensional units so that the  surfaces are at $z^-=-1$ and $z^+=0$. The standard vertical modes are  $\sp_0=1$
and, for $n \geq 1$ 
\beq
\sp_n = \sqrt{2}\cos (n \pi z)\com
\label{std_modes_constN}
\eeq
with $\kappa_n = n \pi$.

We consider the basic state of the Eady problem with streamfunction
\beq
\psi = - \underbrace{(1+z)}_{U} y\com 
\label{eadyStream}
\eeq
and zero interior PV $q=0$ and $\beta=0$. The surface buoyancies are $\vthtb = - y$.

The Galerkin expansion of the PV $q=0$ is exact: $\gq_{\nmax}=0$ and therefore  $\qG_{\nmax}=0$. The truncated Galerkin expansion of $\psi$ follows from  either \eqref{straight9} or  \eqref{sol11} and is
\beq
\psiG_{\nmax} = - \underbrace{\left[ \half \sp_0 + 2 \sqrt{2} \left(\frac{\sp_1}{\pi^2} + \frac{\sp_3}{(3\pi)^2}+ \cdots + \frac{\sp_{\nmax}}{(\nmax\pi)^2}  \right) \right]}_{\UGN} y\per
\label{easyexact2}
\eeq
(We assume that $\nmax$ is odd, so that the last term in the truncated series is as above.) Despite the nonzero derivative of $\psi$ at the boundaries, the series in \eqref{easyexact2} is absolutely and uniformly convergent on the closed interval $-1\leq z\leq 0$. The $\nmax^{-2}$ behavior of the series \eqref{easyexact2}
 ensures uniform convergence, e.g., using the M-test \citep[][]{HN2001}. There are no Gibbs oscillations and a modest number of terms provides a good approximation to the base velocity $U$ (Figure \ref{eady_mean_state}a). 

Now, to illustrate \eqref{unpleasant} and \eqref{Bretherton0}, notice that
\begin{align}
\left(\lap + \sL\right)\psiG_{\nmax} &=  2 \sqrt{2} \left(\sp_1+\sp_3 + \cdots \sp_{\nmax} \right)y \label{easyexact4.1} \\
                                     &= 2\,\frac{\sin\left[(\nmax+1)\pi z\right]}{\sin(\pi z)}\,y\per
\label{easyexact4}
\end{align}
The series \eqref{easyexact4.1} does not converge in a pointwise sense. However, in a distributional sense \cite[][ch.~11]{HN2001}, the exact sum in \eqref{easyexact4} does converge to $\delta$-distributions on the boundaries; see figures \ref{eady_mean_state}(b) and \ref{eady_mean_state}(c). These boundary $\delta$-distributions are the Brethertonian PV sheets \citep{bretherton1966}.

\section{Three approximations} 
\label{sec_three_approx} 
 In  \eqref{unpleasant} we noted that the Galerkin approximations to $\psi$ and $q$ do not exactly satisfy the inversion relation. To address  this error there are at least  three different approximations one can make:

\textbf{Approximation A:} Use the truncated sum  $\psiG_{\nmax}$ in   \eqref{straight8} as a least-squares   Galerkin approximation to the streamfunction $\psi$. But do not   use the Galerkin approximation for $q$.  Instead, \textit{define} the approximate PV, $\qA(x,y,z,t)$, so that the interior inversion relation   is satisfied exactly:
\beq
  \qA \defn \big( \lap +\sL\big) \psiG_{\nmax}\per
  \label{approxA}
 \eeq
This is the approximation introduced by \cite{flierl1978}, which is now regarded as the standard in physical oceanography.  Note that $\qA$ in \eqref{approxA} is not the least-squares approximation to the exact $q$. Moreover, the approximation $\qA$ approaches  the Brethertonian PV on the right of \eqref{Bretherton0} as $\nmax \to \infty$.
 
\textbf{Approximation B:} Use the truncated sum  $\qG_{\nmax}$ in   \eqref{straight13} as a least-squares    Galerkin approximation to the PV  $q$. But do not   use the Galerkin approximation for $\psi$. Instead, \textit{define} the approximate streamfunction, $\psiB(x,y,z,t)$, as the solution to the inversion boundary value problem
  \beq
 \big( \lap +\sL\big)  \psiB = \qG_{\nmax} \com 
 \label{approxB1}
 \eeq
 with boundary conditions
 \beq
\bur \p_z  \psiB= \vartheta^{\pm}\per
\label{approxB2}
 \eeq
This is the approximation introduced  by \cite{tulloch_smith2009b}.
Notice that \eqref{approxB1}  and \eqref{approxB2} is an approximation to the exact inversion problem because the interior source is $\qG_{\nmax}$, rather than $q$. In other words,  $\psiB$ is an exact solution to an approximate  version of the inversion problem. But $\psiB$  is not a least-squares approximation to the exact $\psi$, and nor can $\psiB$  be written as a finite sum of vertical modes.

\textbf{Approximation C:} Use truncated Galerkin approximations $\psiG_{\nmax}$ and $\qG_{\nmax}$ for \textit{both} $\psi$ and $q$. In this case, as indicated in \eqref{unpleasant}, the inversion equation will not be satisfied exactly by the approximate streamfunction and PV. But instead, one will have true least-squares  approximations to both $\psi$ and $q$. To our knowledge approximation C, correctly accounting for the surface-buoyancy boundary terms, has not been previously investigated. 
 
 In approximation A there are $\nmax + 1$ modal amplitudes. In approximations B and C there are $\nmax+3$ degrees of freedom: the $\nmax+1$ modal amplitudes $\gq_n$ and the two surface buoyancy fields $\vth^{\pm}$. The three approximations are equivalent when $\vth^{\pm}=0$. Approximation C is a true Galerkin approximation and it is important to understand its limitations and advantages relative to the better known alternatives A and B.

Once an approximation has been chosen, one needs to construct evolution equations for the Galerkin coefficients using the QG equations \eqref{QGPV1} and \eqref{bc1}. In the next three sub-sections, we derive evolution equations and the associated inviscid conservation laws for the three approximations outlined above. After testing, we recommend C as the most reliable approximation using standard vertical modes.

\subsection{Approximation A}
 Following \cite{flierl1978}, in approximation A the $\nmax$-mode approximate PV is defined  via \eqref{approxA} and, using the modal representation for $\psiG_{\nmax}$ in \eqref{straight8}, this is equivalent to 
 \beq
 \label{invr_A}
 \qA \defn \sum_{n=0}^{\nmax} \helmn \breve{\psi}_n(x,y,t) \, \sp_n(z)\com
 \eeq
 where $\helmn$ is the Helmholtz operator in \eqref{helmdef}.
 Following the appendix of \cite{flierl1978}, one can use Galerkin projection of the nonlinear evolution equation  \eqref{QGPV1} onto the modes $\sp_n$ to obtain $\nmax+1$ evolution equations for the coefficients $\gpsi_n$:
 \begin{align}
\p_t \helmn \gpsi_n + \sum_{m=0}^{\nmax} \sum_{s=0}^{\nmax}   \Xi_{nms} \,  \sJ\left(\gpsi_m, \helms \gpsi_s\right) 
 + \beta \p_x \gpsi_n=0\com
\label{modeFl1}
\end{align}
where
\beq
\Xi_{nms} \defn \tfrac{1}{h} \int_{z^-}^{z^+} \!\! \sp_n\sp_m \sp_s \, \dd z \per
\label{Xidef}
\eeq
Note that $\Xi_{nms}$ cannot be computed exactly except in cases with simple buoyancy frequency profiles. But it suffices to compute $\Xi_{nms}$ to high accuracy, e.g. using Gaussian quadrature.

\cite{flierl1978} implicitly assumed that $\vth^+=\vth^-=0$, so that the surface terms in \eqref{sol11} vanish and then there is no difference between $\qA$ and $\qG_{\nmax}$. But in general, with nonzero surface buoyancy, we can append evolution equations for $\vtht$ and $\vthb$ to approximation A. That is, in addition to the $\nmax+1$ modal equations in \eqref{modeFl1}, we also have 
\beq
\p_t \vth^{\pm} + \sum_{n=0}^{\nmax} \sp_n^{\pm} \, \sJ(\gpsi_n,\vth^{\pm})=0\per
\label{BC7}
\eeq
Above we have evaluated the $\psi$-series \eqref{straight8} at $z^{\pm}$ to approximate  $\psi^{\pm}$ in the surface boundary conditions. This approach is not satisfactory because the resulting surface buoyancy equations \eqref{BC7} are dynamically passive i.e., $\vth^+$ and $\vth^-$ do not affect the interior evolution equations in \eqref{modeFl1}. 

\subsubsection*{ Quadratic conservation laws}

Appendix A shows that Approximation A has the energy conservation 
\beq
\label{cons_E_A}
\frac{\dd}{\dd t} \sum_{n=0}^{\nmax}  \int \halfrho  (\nabla \breve{\psi}_n)^2   + \halfrho  \kappa_n^2  \breve{\psi}_n^2\, \,\dd S = 0\per
\eeq
To obtain the energy analogous to $E$ in \eqref{E_defn}, the modal sum above is  multiplied by the depth $h$.
With $\beta\neq0$, approximation A has the potential enstrophy conservation law,
\beq
\label{cons_Ens_A}
\frac{\dd}{\dd t}\sum_{n=0}^{\nmax} \int  \half \big(\helmn \breve \psi_n \big)^2 \dd S = 0\per
\eeq
With $\beta \neq 0$, the analog of the exact potential enstrophy \eqref{ensdens} is not conserved.
Finally, with the surface equations in \eqref{BC7}, approximation A also conserves surface buoyancy variance  as in \eqref{sB_cons}.

\subsection{Approximation B}

Approximation B  begins with the observation  that  the exact solution of the inversion problem in \eqref{inv3} and \eqref{inv5} can be decomposed as
\beq
\psi =\phi + \sigma
\label{Bdecomp}
\eeq
where $\phi(x,y,z,t)$ is the ``interior streamfunction" and $\sigma(x,y,z,t)$ is the ``surface streamfunction" \citep{lapeyre_klein2006,tulloch_smith2009b}.

The surface streamfunction  $\sigma(x,y,z,t)$ is defined as the solution of the  boundary value problem
\beq
  \label{inv_sigma}
  \left(\lap +\sL\right) \sigma = 0\com
\eeq
  with inhomogeneous Neumann boundary conditions 
\beq
  \label{bc_sigma}
  \bur \p_z \sigma\left(z^\pm\right) = \vartheta^{\pm} \per 
\eeq
In approximation B, one  must always solve for the surface streamfunction using methods other than a truncated series. The solution of the surface problem  \eqref{inv_sigma} and \eqref{bc_sigma} in  terms of standard vertical modes is
\beq
\sigma = \sum_{n=0}^\infty \gsigma_n(x,y,t)\sp_n(z)\com \quad \text{and} \quad \gsigma_n = \tfrac{1}{h}\int_{z^{-}}^{z^{+}} \!\!\!\!  \sp_n \sigma \, \dd z\per
\label{resort1}
\eeq
where
\beq
\label{gsigma_defn}
\helmn \gsigma_n = -\tfrac{1}{h} (\sp_n^{+}\vtht - \sp_n^{-}\vthb) \com
\eeq
and $\helmn$ is the $n$'th mode  Helmholtz operator defined in \eqref{helmdef}. In  \eqref{resort1} and \eqref{gsigma_defn} we have a solution for the surface streamfunction $\sigma$, with nonzero vertical derivative $\sigma_z$ at the surfaces, in terms of vertical modes with zero derivative. Truncations of the series \eqref{resort1} behave similarly to the truncated series \eqref{easyexact2}:  convergence is absolute and uniform (see appendix A).

The interior streamfunction $\phi(x,y,z,t)$ is defined as the solution of the  boundary value problem
\beq
  \label{inv_phi}
  \left(\lap +\sL\right) \phi = q\com
  \eeq
with homogeneous Neumann boundary conditions 
 \beq
  \label{bc_phi}
  \bur \p_z \phi\left(z^\pm\right) = 0\per 
  \eeq
  
  Approximation B assumes that one can solve the surface problem in  \eqref{inv_sigma} and  \eqref{bc_sigma} without resorting to truncated versions of the series in \eqref{resort1}. For instance, with constant or exponential stratifications one can find closed-form, exact expressions for $\sigma$ \citep{tulloch_smith2009b,lacasce2012}. In particular, approximation B requires that  the two unknown Dirichlet boundary-condition functions $\sigma^{\pm}=\sigma(z^{\pm})$ can be obtained efficiently from specified Neumann boundary-condition functions $\vtht$ and $\vthb$.  The Eady problem, discussed below in section \ref{eadysec}, is a prime example in which one can obtain this Neumann-to-Dirichlet map.

Once $\sigma$ is in hand, the approximate streamfunction is
 \beq
 \psiB =\phiB + \sigma \com
 \label{decomp1}
 \eeq
 where $\phiB(x,y,z,t)$ is obtained by solving the interior  inversion problem \eqref{inv_phi} with the right hand side replaced by the Galerkin approximation $ \qG_{\nmax}$ defined in \eqref{straight13} and \eqref{straight17}. The exact solution of this approximation to the interior inversion problem is 
\beq
\phiB = \sum_{n=0}^{\nmax} \gphi_n(x,y,t)\sp_n(z)\com
\eeq
where
\beq
\label{qbreve_B}
\gphi_n \defn \tfrac{1}{h} \int_{z^-}^{z^+} \!\!\!\! \sp_n \phi \, \dd z\com
\qquad \text{and} \qquad \helmn \gphi_n = \gq_n \per
\eeq 

To obtain the  approximation B  evolution equations we introduce the streamfunction \eqref{decomp1} into the QGPV equation  \eqref{QGPV1} and project onto mode $n$ to obtain
\begin{align}
    \p_t \helmn \breve \phi_n  + \sum_{m=0}^{\nmax} & \sum_{s=0}^{\nmax}  \Xi_{nms} \,  \sJ\left(\gphi_m, \helms \gphi_s\right) + \beta  \p_x \left( \gphi_n + \breve \sigma_n\right) \nonumber \\ 
  & + \sum_{s=0}^{\nmax}  \tfrac{1}{h} \int_{\zm}^{\zp}\!\!\!\!\sp_n \sp_s \sJ\left(\sigma \com\helms \gphi_s\right) \, \dd z      = 0\com
\label{modeEqB}
\end{align}
with $\Xi_{nms}$ defined in \eqref{Xidef}. Approximation B assumes that the remaining integral on the second line  of \eqref{modeEqB} can be evaluated exactly. This is only possible for particular models of the $N(z)$ (e.g., constant buoyancy-frequency profiles). In practice, however, it may suffice to compute the integral  on the second line \eqref{modeEqB} very accurately, e.g. using Gaussian quadrature.

The evolution equations for approximation B are completed with the addition of buoyancy-advection at the surfaces 
\beq
    \p_t \vth^{\pm} + \sJ(\sigma^{\pm},\vth^{\pm})  + \sum_{n=0}^{\nmax}  \,\sp_n^{\pm}  \sJ( \breve  \phi_n,\vth^{\pm}) =0\per
\label{BC_B}
\eeq
With \eqref{modeEqB} and \eqref{BC_B} we have  $\nmax +3$ evolution equations for the $\nmax + 3$ fields $\gphi_0, \gphi_1, \cdots \gphi_{\nmax}$ and $\vartheta^{\pm}$.

\subsubsection*{Quadratic conservation laws }
Approximation B  conserves surface buoyancy variance. But the  conservation laws for energy and potential enstrophy are problematic. The analog of the exact total energy \eqref{E_defn} is not generally conserved in approximation B (Appendix A). With $\beta=0$, approximation B has a potential enstrophy conservation
\beq
\label{cons_Ens_C}
\frac{\dd}{\dd t} \sum_{n=0}^{\nmax}  \int\half (\helmn \breve  \phi_n)^2 \dd S = 0\per
\eeq
But with $\beta\neq 0$  the analog of the exact potential enstrophy \eqref{ensdens} is not conserved (Appendix A).

\subsection{Approximation C}

Because method C  approximates \textit{both} the streamfunction and the  PV by  Galerkin series, the derivation of the modal equations is very straightforward compared with the calculations in appendix A of Flierl (1978): one simply substitutes the truncated Galerkin series for the streamfunction \eqref{straight8} and  PV \eqref{straight13} into the QGPV equation \eqref{QGPV1}, and then projects onto mode $n$ to obtain
\beq
\p_t \gq_n  + \sum_{m=0}^{\nmax} \sum_{s=0}^{\nmax}  \Xi_{nms} \,  \sJ\left(\gpsi_m, \gq_s\right) + \beta \p_x \gpsi_n=0\com
\label{modeEqC}
\eeq
where $\Xi_{nms}$ is defined in \eqref{Xidef}, and  we recall the relation between $\gpsi_n$ and $\gq_n$ from \eqref{sol11}
\beq
\gq_n =  \helmn \gpsi_n+\tfrac{1}{h}   \left( \sp_n^+\,  \vtht  -  \sp_n^-\, \vthb \right) \per
 \label{sol111}
 \eeq
 In approximation C there are $\nmax + 3$ degrees of freedom: the $\nmax+1$ modal amplitudes $\gpsi_n$ and the two surface buoyancy fields $\vth^{\pm}$. The approximation $C$ evolution equations are completed by advection of the  surface buoyancy 
 \beq
\p_t \vth^{\pm} + \sum_{n=0}^{\nmax} \sp_n^{\pm} \, \sJ(\gpsi_n,\vth^{\pm})=0\per
\label{BC77}
\eeq
We emphasize that  in  approximation C the surface buoyancy  fields $\vthtb$ are not passive: $\gpsi_n$, $\gq_n$, and $\vthtb$ are related through \eqref{sol111}. 

Finally, note that approximation C is  recovered from approximation B if the surface streamfunction is represented by a truncated version of the series \eqref{resort1}.

\subsubsection*{Quadratic conservation laws}

Approximation C conserves surface buoyancy variance as in  \eqref{sB_cons}. Total energy is also conserved

\beq
\label{cons_E_C}
\frac{\dd}{\dd t}\sum_{n=0}^{\nmax} \int   \halfrho  |\nabla \breve{\psi}_n|^2 + \halfrho   \kappa_n^2  \breve{\psi}_n^2 \,  \dd S  = 0\per
\eeq
As in approximation B, conservation of potential enstrophy is troublesome. With $\beta = 0$, approximation C has a potential enstrophy conservation law
\beq
\label{cons_Ens_C_beta}
\frac{\dd}{\dd t} \sum_{n=0}^{\nmax}  \int \half \breve{q}_n^2 \, \dd S = 0\per
\eeq
But with $\beta\neq 0$, approximation C does not conserve the analog of the exact potential enstrophy \eqref{ensdens} (Appendix A).

\section{The Eady problem \label{eadysec}}
We use classical linear stability problems with nonzero surface buoyancy to illustrate how solutions to specific problems can be constructed and to assess the relative merit and efficiency of approximations A, B, and C. The linear analysis does not provide the full picture of convergence of the approximate solutions. Nonetheless, in turbulence simulations forced by baroclinic instability, it is necessary (but not sufficient) to accurately capture the linear stability properties.

We use nondimensional variables so that the surfaces are at $z^{+}=0$ and $z^{-}=-1$. The Eady exact base-state velocity is given by \eqref{eadyStream} with zero PV $q=0$ and $\beta=0$. 

\subsection{Approximation A}

\begin{figure*}
\begin{center}
\includegraphics[width=12pc,angle=0]{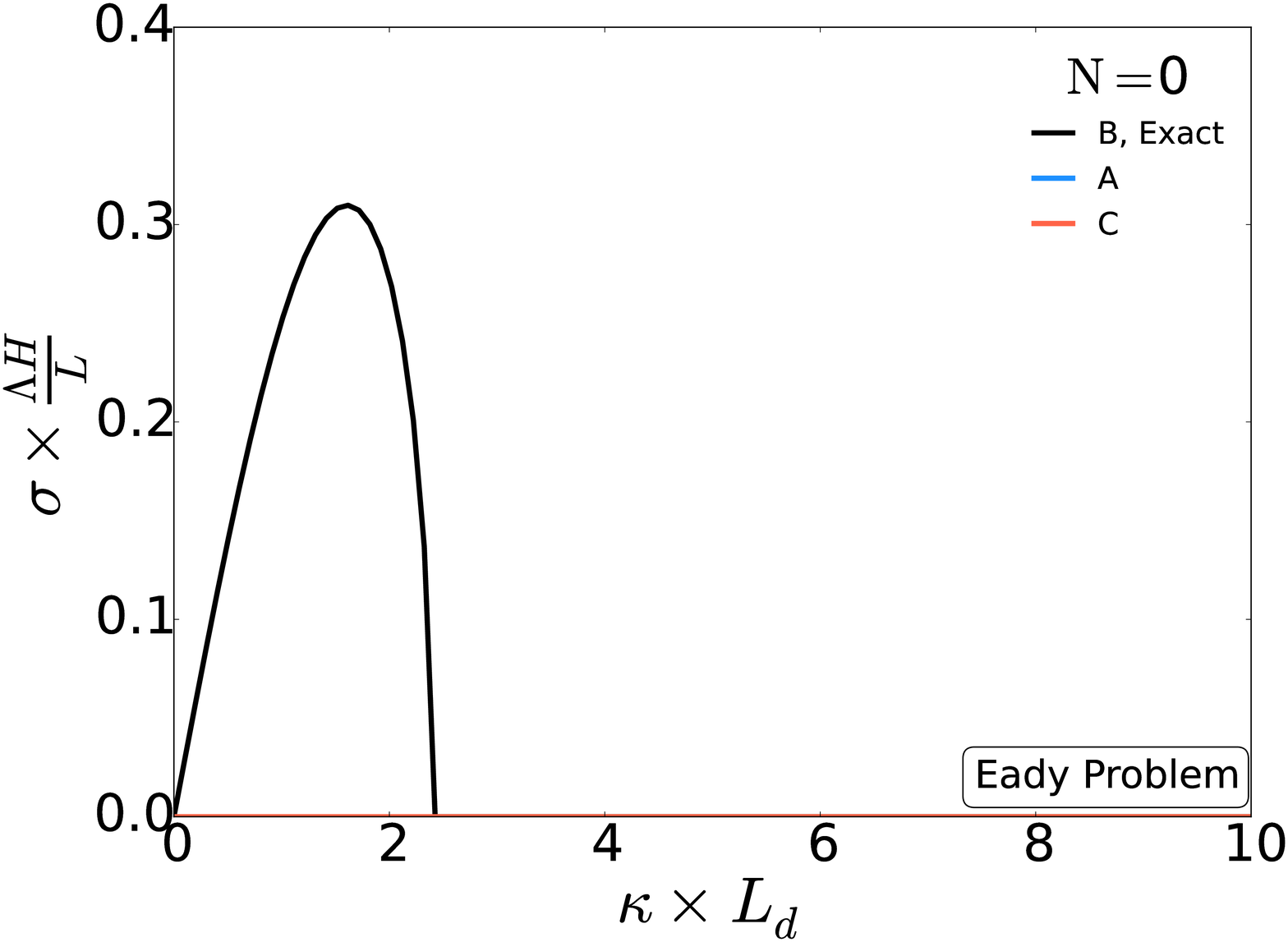}\includegraphics[width=12pc,angle=0]{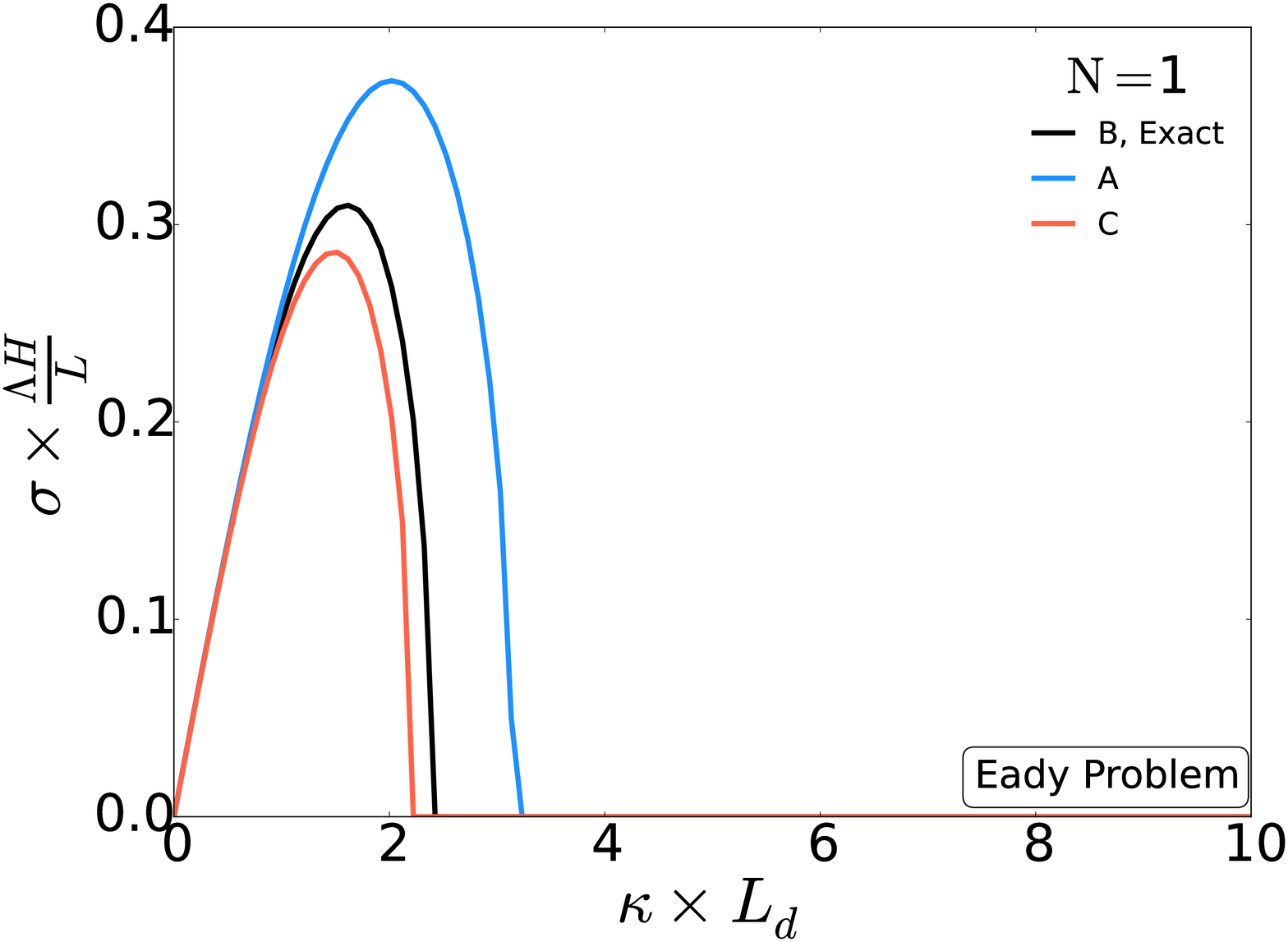} 
\includegraphics[width=12pc,angle=0]{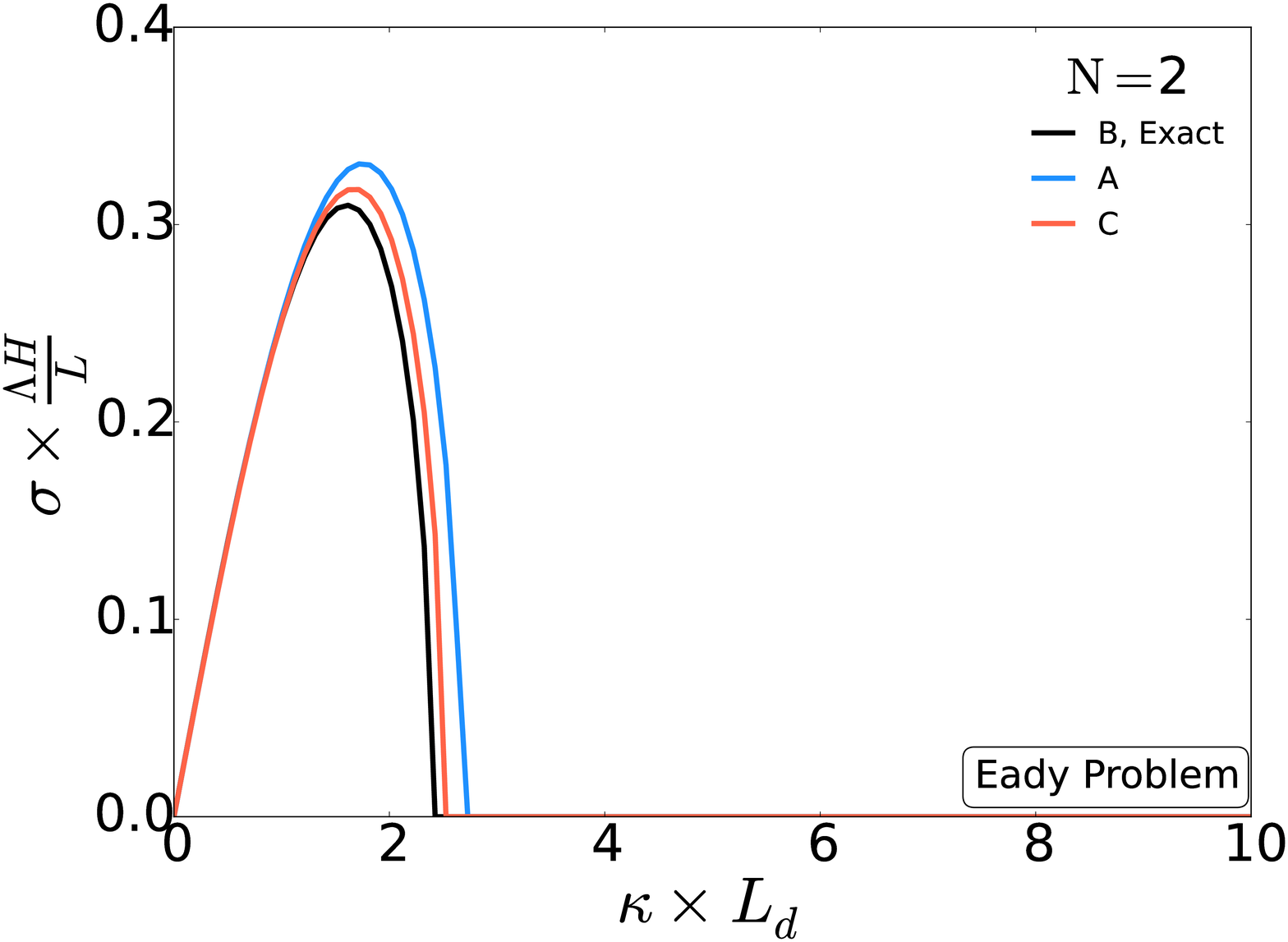}\\\includegraphics[width=12pc,angle=0]{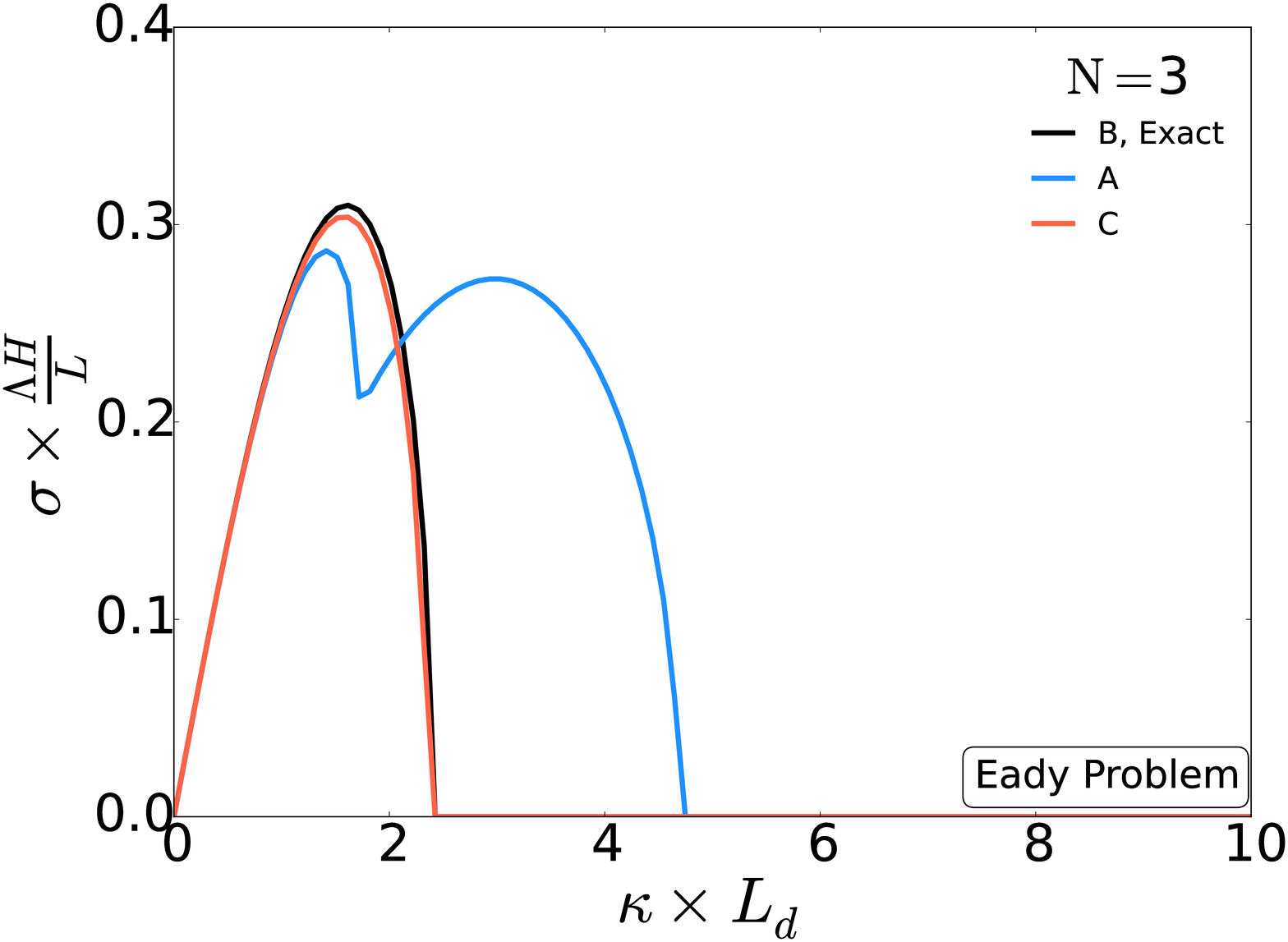} 
\includegraphics[width=12pc,angle=0]{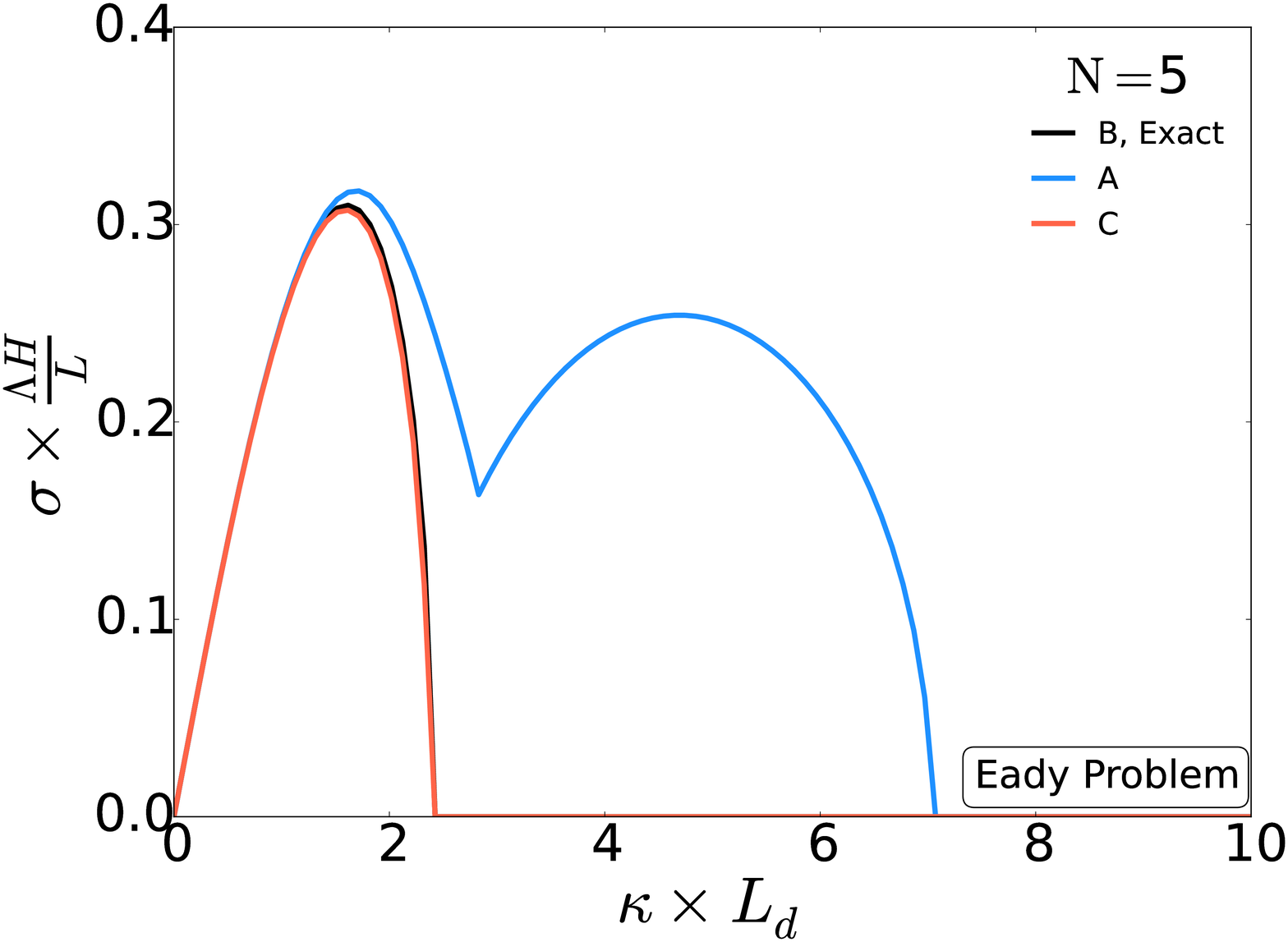}\includegraphics[width=12pc,angle=0]{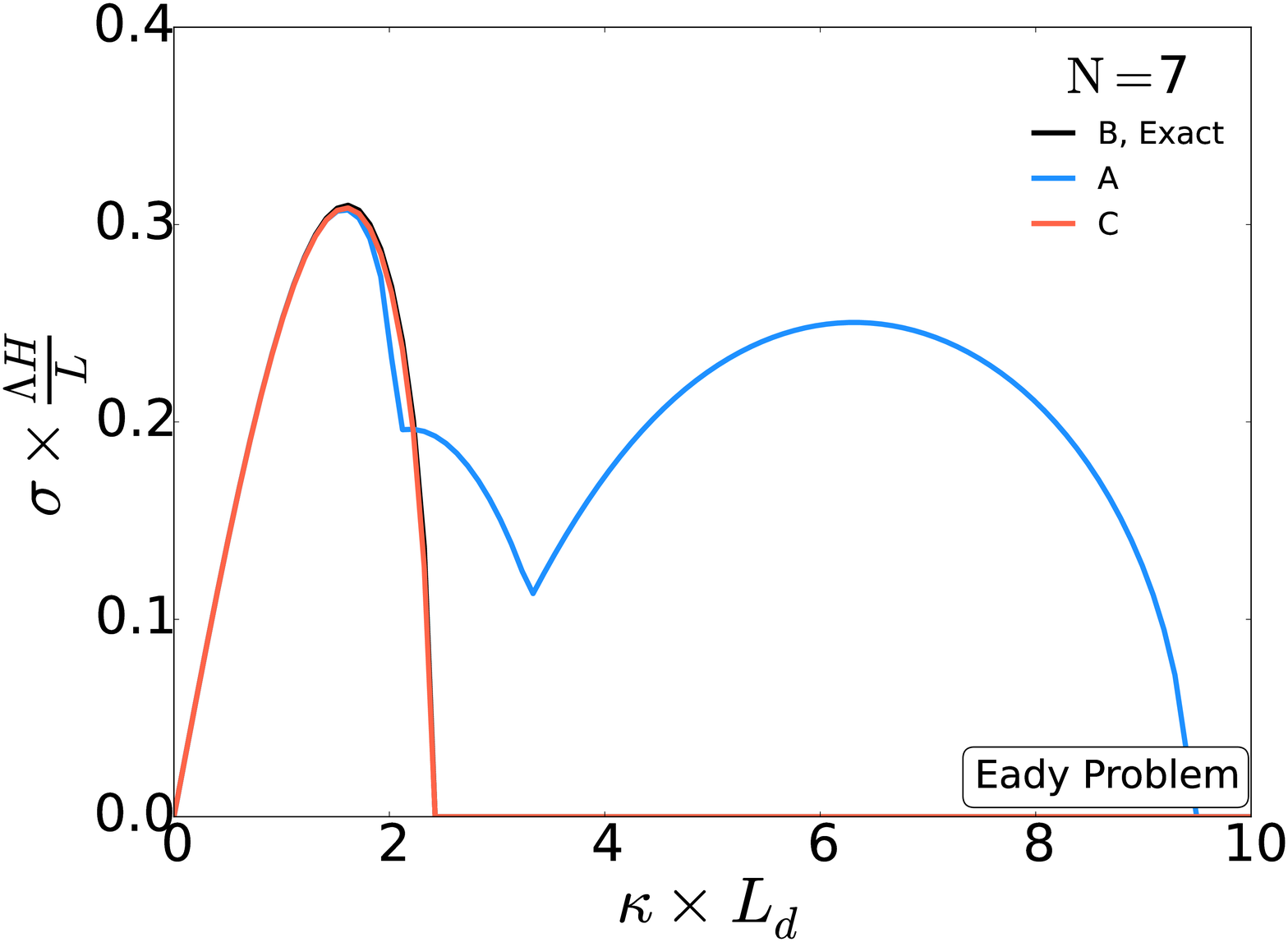} \\
\includegraphics[width=12pc,angle=0]{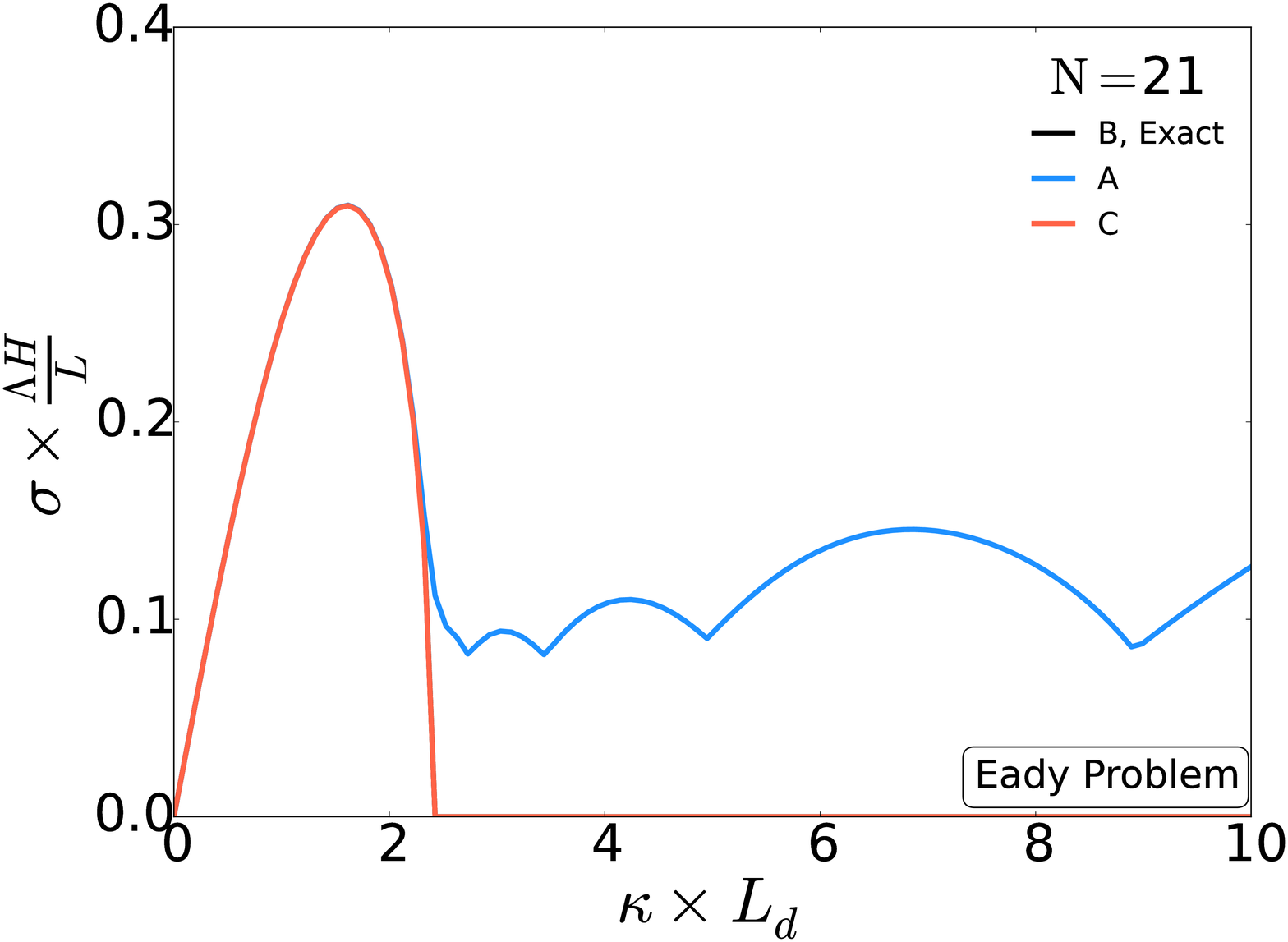}\includegraphics[width=12pc,angle=0]{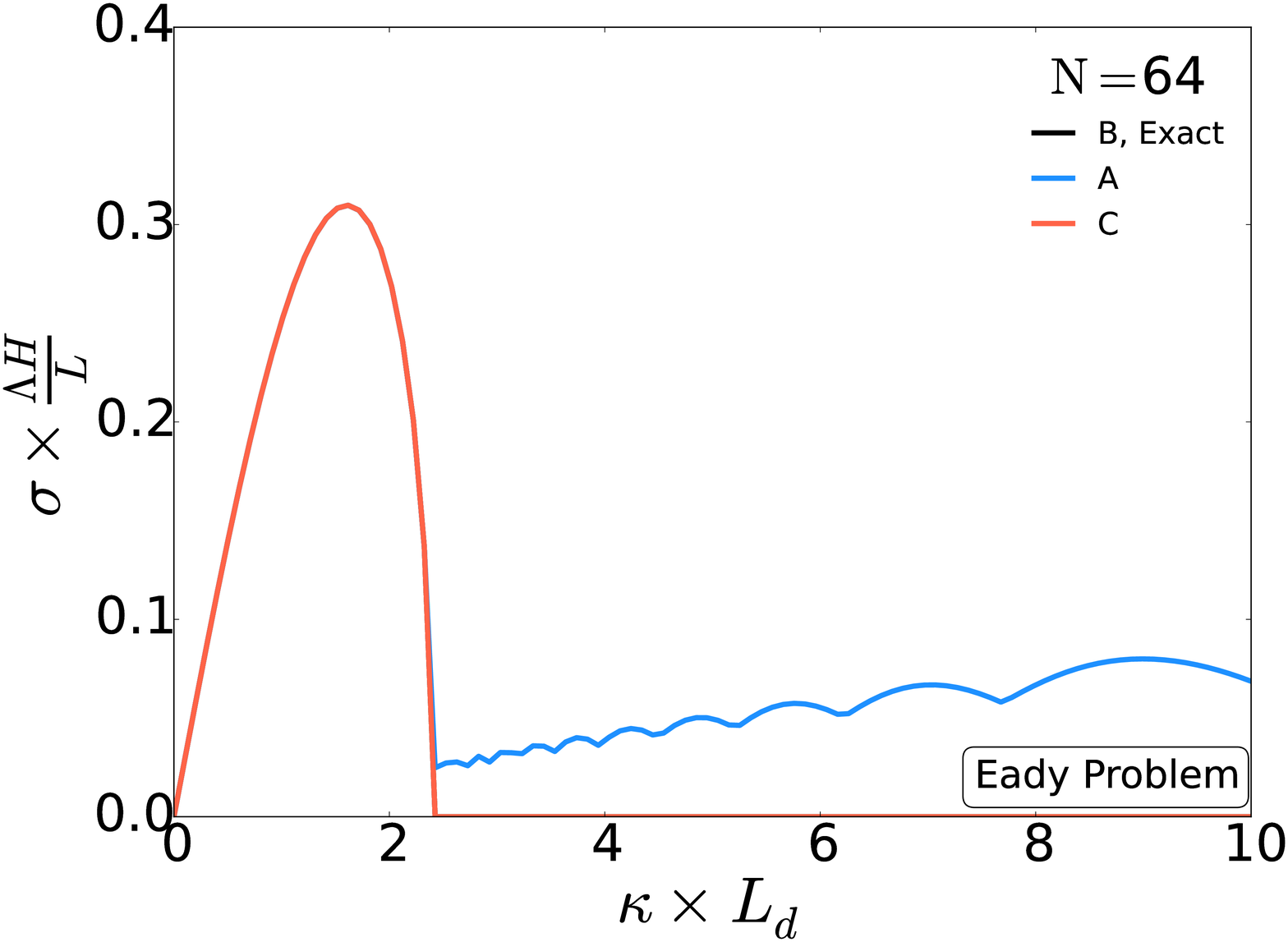}\includegraphics[width=12pc,angle=0]{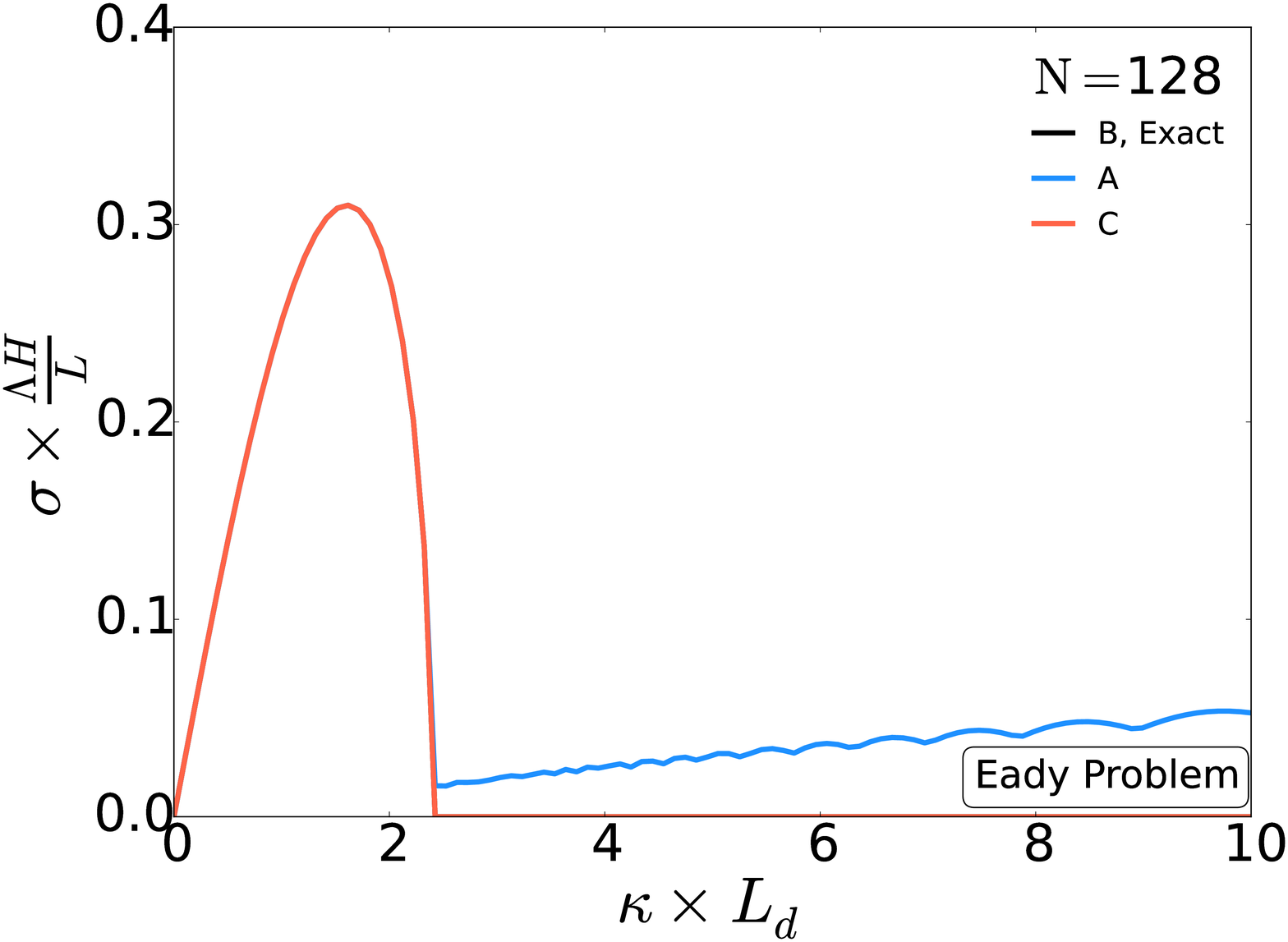}
\end{center}
\caption{Growth rate for the Eady problem as a function of the zonal wavenumber ($l=0$) using approximations A, B (exact), C with various number of baroclinic modes ($\nmax$).}
\label{eady_growth_rate}
\end{figure*}

While the surface fields $\vthtb$ are dynamically passive in approximation A, the Eady problem can still be considered because the base-state PV defined via \eqref{invr_A} converges to $\delta$-distributions on the boundaries (Section \ref{galerk_sec}). 

The base-state velocity in Approximation A is given by the series \eqref{easyexact2} and is a good approximation to the exact base-state velocity \eqref{eadyStream}.  But, according to approximation A, there is  a nonzero interior base-state PV gradient given by the series \eqref{easyexact4}. As $\nmax \to \infty$ the PV gradient in \eqref{easyexact4} converges in a distributional sense to Brethertonian  sheets at $z=0$ and $-1$.   But for numerical implementation of approximation A we stop short of $\nmax = \infty$. While the PV gradient is much larger at the boundaries, there is always interior structure in the PV (Figure \ref{eady_mean_state}c). We show that this spurious  interior PV gradient has a strong and unpleasant effect on the approximate solution of the Eady stability problem.

To solve the Eady linear stability we linearize the interior equations \eqref{modeFl1} about the base-state velocity  in \eqref{easyexact2} and the PV gradient in \eqref{easyexact4}. We assume ${\gq}_k = \hat{q}_k \exp[{\ii(k \, x + l \, y  - \omega^A\,t)}]$, etc, to obtain a $(\nmax+1)\times(\nmax+1)$ eigenproblem
\beq
\label{eigEadyA}
\sum_{m=0}^{\nmax}\sum_{s=0}^{\nmax} \Xi_{nms} \Big[ \gU_m \, \hat q_s +  \p_y \, \gQ_s \,\hat{\psi}_m \Big] = c^A \hat q_n,
\eeq
where $\gQ_s$ are the coefficients of the series \eqref{easyexact4} and $c^A \defn \omega^A/k$. The eigenproblem \eqref{eigEadyA} can be recast in the matrix form $\sA \sq = c^A \sq$, where $\tilde\sq=[\hat{q}_0,\hat{q}_1,\ldots,\hat{q}_{\nmax-1},\hat{q}_{\nmax}]^\sT$ (Appendix B) and solved with standard methods.

\begin{figure}[ht]
\begin{center}
\includegraphics[width=19pc,angle=0]{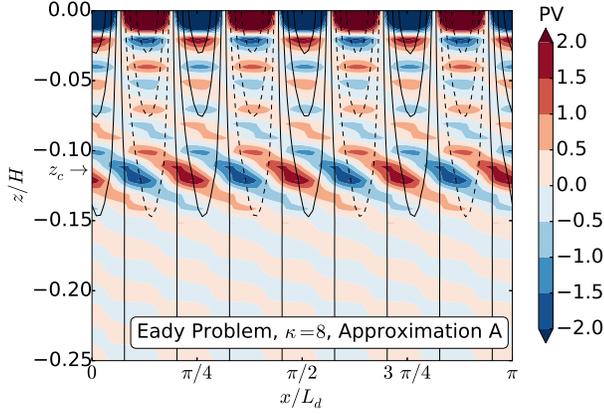}
\end{center}
\caption{Structure of $\kappa = 8$ unstable mode for the Eady problem obtained using approximation A and $\nmax=64$. Streamfunction  is the black curves and  PV is the colors. The streamfunction slightly tilts westward as $z$ increases. One can see the unphysical critical layer associated with the fast-oscillating base-state PV. The critical level, $z_c$, is the depth where the unstable wave speed matches the velocity of the base-state. Only the top quarter of the domain is shown.}
\label{eady_wave_structure}
\end{figure}

 Figure \ref{eady_growth_rate} shows the growth rate of the Eady instability according to approximation A, and compares this with the exact Eady growth rate. Approximation A does not do well, especially at large wavenumbers. The exact Eady growth rate has a high-wavenumber cut-off. At moderate values of $\nmax$, such as $3$, $5$ and $7$ approximation A produces unstable ``bubbles" of instability at wavenumbers greater than the high-wavenumber cut-off. The  growth rates in these bubbles are comparable to the true maximum growth rate. As $\nmax$ increases, the unstable bubbles are replaced by a long tail of unstable modes with a growth rate that slowly increases with  $\kappa$. These spurious high-wavenumber instabilities  are due to the rapidly oscillatory interior PV gradient  which supports  unphysical  critical layers: see   Figure \ref{eady_wave_structure}.  

\subsection{Approximation B, the exact solution}
In approximation B, the zero PV in the Eady problem implies $\breve q_n = \gphi_n = 0$. The $\nmax+1$ modal equations (with $\beta = 0$) are trivially satisfied; there is no interior contribution $(\phiB = 0)$. Thus approximation B solves the Eady problem exactly. 
    
Assuming $\sigma = \hat{\sigma}(z) \exp[\ii(k \,x + l \,y - \omega^B \, t)]$, we obtain the solution to the surface streamfunction inversion problem \eqref{inv_sigma}-\eqref{bc_sigma}
\beq
\hat{\sigma}(z) = \frac{\cosh[\kappa(z+1)]}{\kappa\sinh\kappa} \, \vtht -\frac{\cosh(\kappa z)}{\kappa\sinh\kappa} \, \vthb \com \label{sigma_eady}
\eeq
where the magnitude of the wavenumber vector is $\kappa = \sqrt{k^2 + l^2}$. We evaluate the surface streamfunction \eqref{sigma_eady} at the boundaries to find the relationship between the streamfunction at the surfaces $\hat{\sigma}^{\pm}$ and the boundary fields $\vthtb$:
\beq
\label{transfer_func}
\begin{bmatrix*} 
\hat{\sigma}^+\\
\hat{\sigma}^-
\end{bmatrix*} 
= \frac{1}{\kappa}
\begin{bmatrix*} 
\coth\kappa  & -\cosech\kappa \\
    \cosech\kappa & -\coth\kappa
\end{bmatrix*}
\begin{bmatrix*} 
\vththat\\
\vthbhat
\end{bmatrix*},
\eeq
The nondimensional linearized boundary conditions \eqref{BC_B} are 
\beq
\label{bc_eady_top_B}
\vththat - \hat{\psi}^+ = c^B \vththat\com \qquad \text{and} \qquad 
-\hat{\psi}^- = c^B \vthbhat\com
\eeq
where $c^B = \omega^B/k$. Using the boundary conditions \eqref{bc_eady_top_B}  in \eqref{transfer_func}
  we obtain an eigenvalue problem
\beq
\label{Bdefn}
\underbrace{ \frac{1}{\kappa} 
\begin{bmatrix*} 
    \kappa - \coth\kappa & \cosech\kappa \\
    -\cosech\kappa  & \coth\kappa \\
\end{bmatrix*}}_{\defn \sB} 
\begin{bmatrix*} 
\vththat\\
\vthbhat
\end{bmatrix*} = c^B \begin{bmatrix*} 
\vththat\\
\vthbhat
\end{bmatrix*}\per
\eeq
The eigenvalues of $\sB$ are given by the celebrated dispersion relation for the Eady problem  \citep{pedlosky1987,vallis2006}
\beq
\label{disp_rel}
c^B = \tfrac{1}{2} \pm \tfrac{1}{\kappa}\Big[ \Big(\tfrac{\kappa}{2}-\tanh\tfrac{\kappa}{2}\Big)\Big(\tfrac{\kappa}{2}-\coth\tfrac{\kappa}{2}\Big) \Big]^{1/2}\per
\eeq

\subsection{Approximation C}

Approximation C expands both the streamfunction and the PV  in standard vertical modes. Thus in the Eady problem the PV is exactly zero, as it should be: $q = \breve{q}_n = 0$.  (This contrasts with approximation A, in which  the differentiation of the truncated series  approximation to the streamfunction induces an unphysical oscillatory base-state PV gradient.) Thus approximation C  does not have the  spurious critical layers that bedevil A.    Moreover, in approximation C, the $\nmax+1$ modal equations (with $\beta=0$) in \eqref{modeEqC} are trivially satisfied, and the inversion relationship \eqref{sol111} provides a simple connection between the  streamfunction and the  fields $\vth^{\pm}$. The base velocity for the Eady problem in approximation C is the series in \eqref{easyexact2} (the same as  A). From the exact shear at the boundaries we obtain the exact base-state boundary variables
\beq
\label{base_state_C}
{\vThtb}= -y\per
\eeq
We linearize the boundary equations \eqref{BC77} about the  base-state  \eqref{easyexact4} and \eqref{base_state_C}, to obtain 
\beq
\label{bc_linear_C}
\partial_t \vthtb + {\UGN}^{\pm}\partial_x \vthtb - \,\sum_{k=0}^{\nmax} \partial_x \breve{\psi}_k \sp_k^{\pm}   = 0 \per
\eeq
Assuming $\vthtb = \vthtbhat\exp[\ii(k \,x + l \,y - \omega^C \, t)]$, and using the inversion relationship \eqref{sol111}, we obtain a $2\times 2$ eigenproblem
\beq
\label{MC}
\sC\,\begin{bmatrix*} 
\vththat\\
\vthbhat
\end{bmatrix*}  = c^C \,\begin{bmatrix*} 
\vththat\\
\vthbhat
\end{bmatrix*} \com
\eeq
where matrix $\sC$ is defined in  appendix C. It is straightforward to show that $c^C$ converges to the exact eigenspeed. i.e., $c^C \to c^B$ as $\nmax \to \infty$ (Appendix B). Figure \ref{eady_growth_rate} shows that approximation C successfully captures the structure of the Eady growth rate even with modest values of $\nmax$.

\subsection{Remarks on convergence}

\begin{figure}[ht]
\begin{center}
    \includegraphics[width=18pc,angle=0]{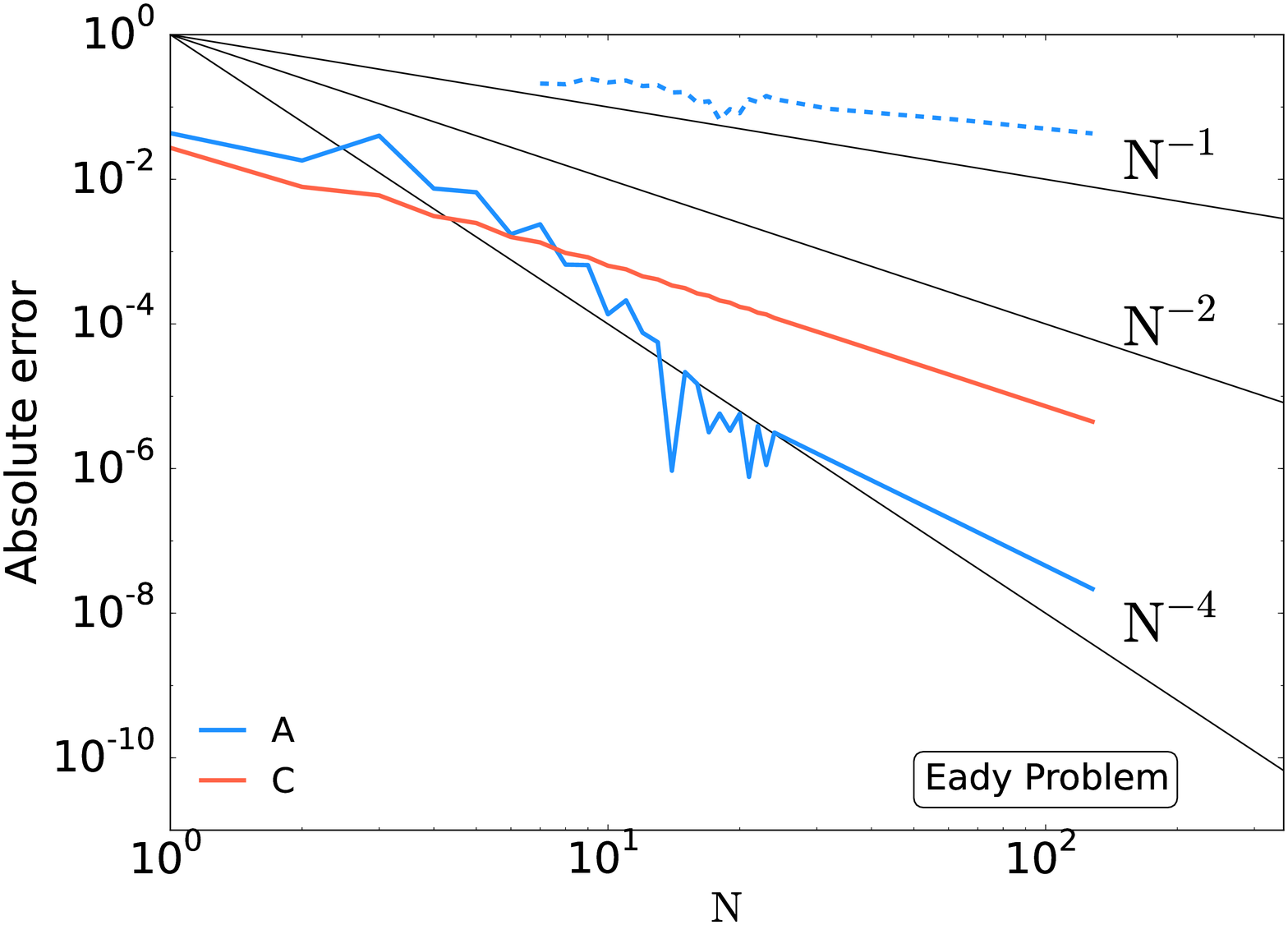}
\end{center}
\caption{Absolute error as a function of number of baroclinic modes ($\nmax$) for the growth rates of the Eady problem. The solid lines show  the error at the exact fastest growing mode ($\kappa \approx 1.6$). The dashed line is the approximation A  error at $\kappa = 8$.}
\label{error_analysis}
\end{figure}

The crudest truncation (i.e. $\nmax=0$) is stable for both approximations A and C (Figure \ref{eady_growth_rate}). With one baroclinic mode ($\nmax=1$) the growth rates $\left(\omega_i =  k \times \mathrm{Im}\{c\}\right)$ are qualitatively consistent with the exact solution, and the results improve with $\nmax=2$. With a moderate number of baroclinic modes modes ($\nmax > 2$) approximations A and C converge rapidly to the exact growth rate at  wavenumbers less than about $2.2$ --- see figure \ref{eady_growth_rate}. But surprisingly the convergence of the growth rate at the most unstable mode ($\kappa \approx 1.6$) is faster in approximation A ($\sim\!\nmax^{-4}$) than in approximation C ($\sim\!\nmax^{-2}$) --- see figure \ref{error_analysis}. However, the convergence in approximation C is uniform: there are no spurious high-wavenumber instabilities.

Figure \ref{error_analysis} also shows that the approximation A convergence of the growth rate to zero  at $\kappa=8$ is slow ($\sim\!\nmax^{-1}$). While the growth rate does  converge to zero at a fixed wavenumber, such as $\kappa=8$, we conjecture that there are always faster growing modes at larger wavenumbers.

\section{The Green problem}
\label{beta_eady_sec}

\begin{figure*}
\begin{center}
\includegraphics[width=12pc,angle=0]{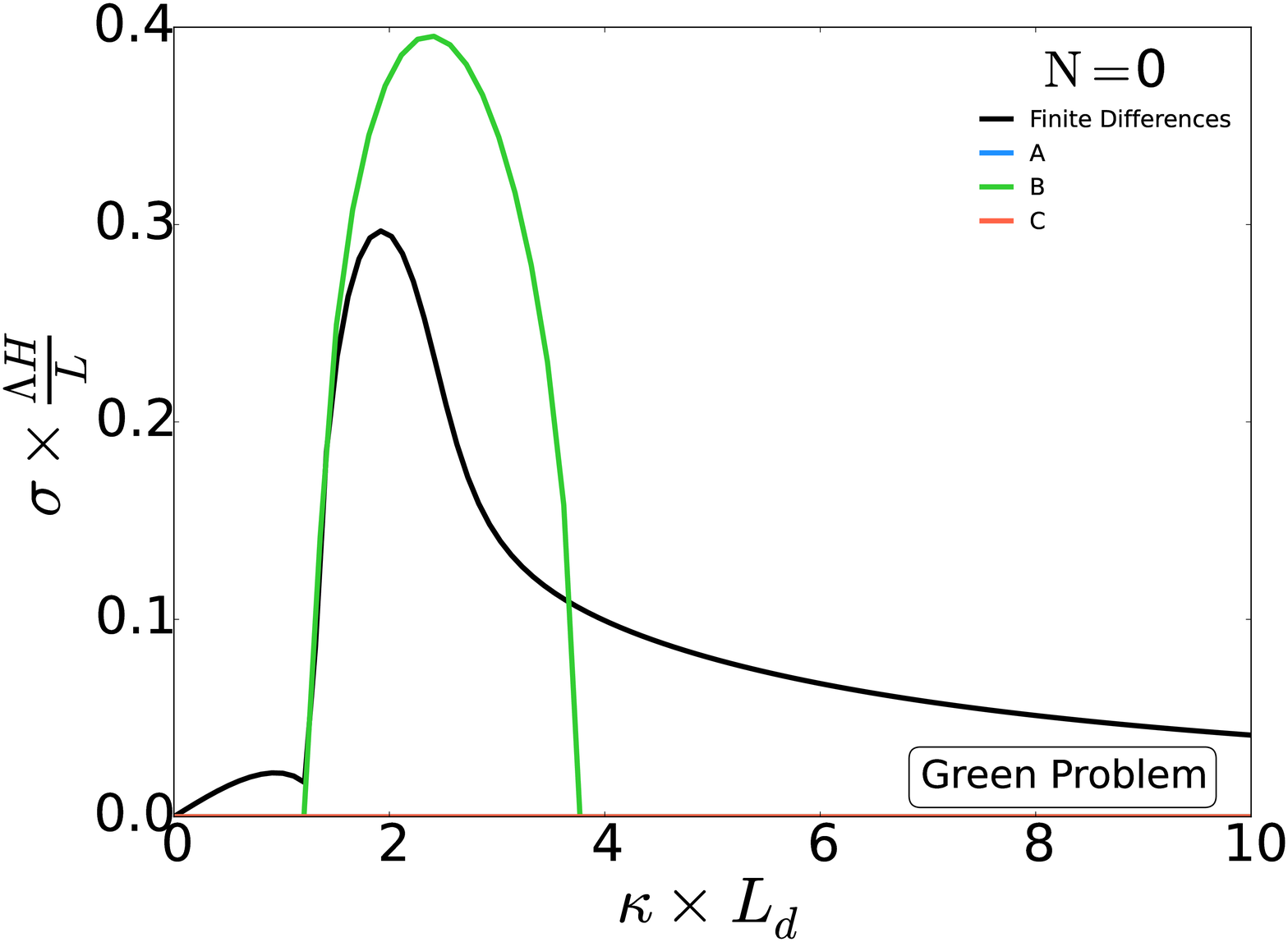}\includegraphics[width=12pc,angle=0]{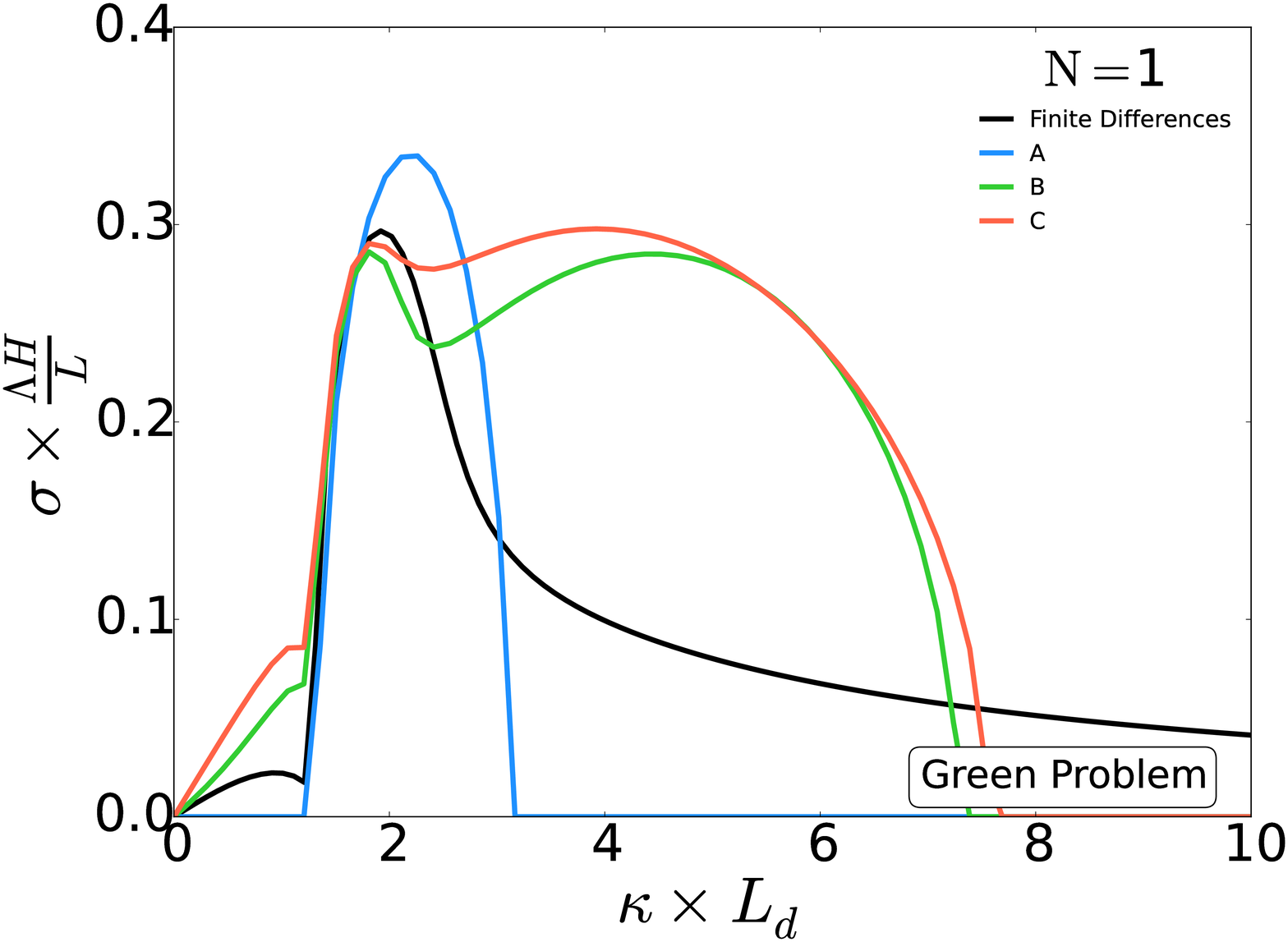} 
\includegraphics[width=12pc,angle=0]{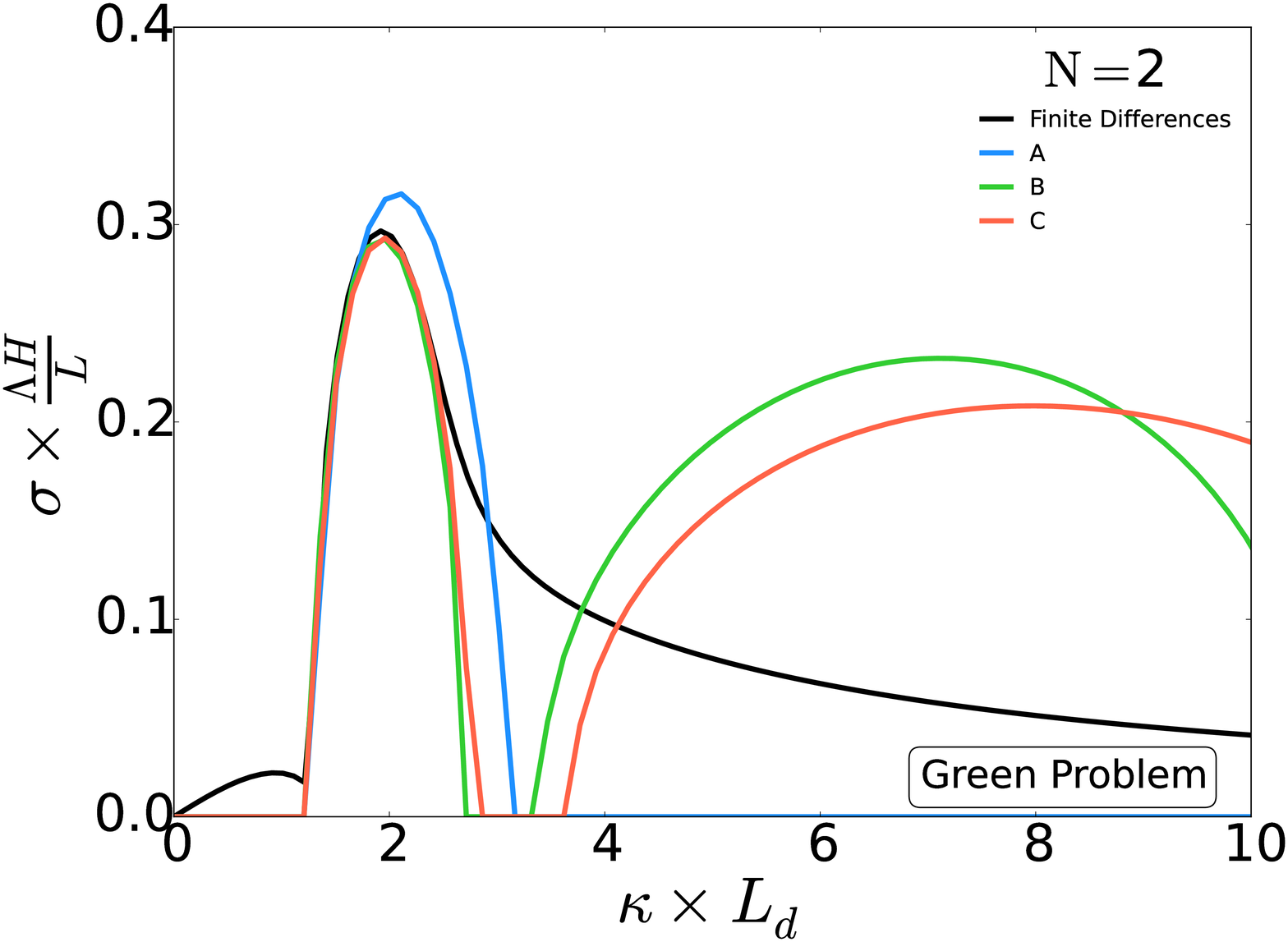}\\\includegraphics[width=12pc,angle=0]{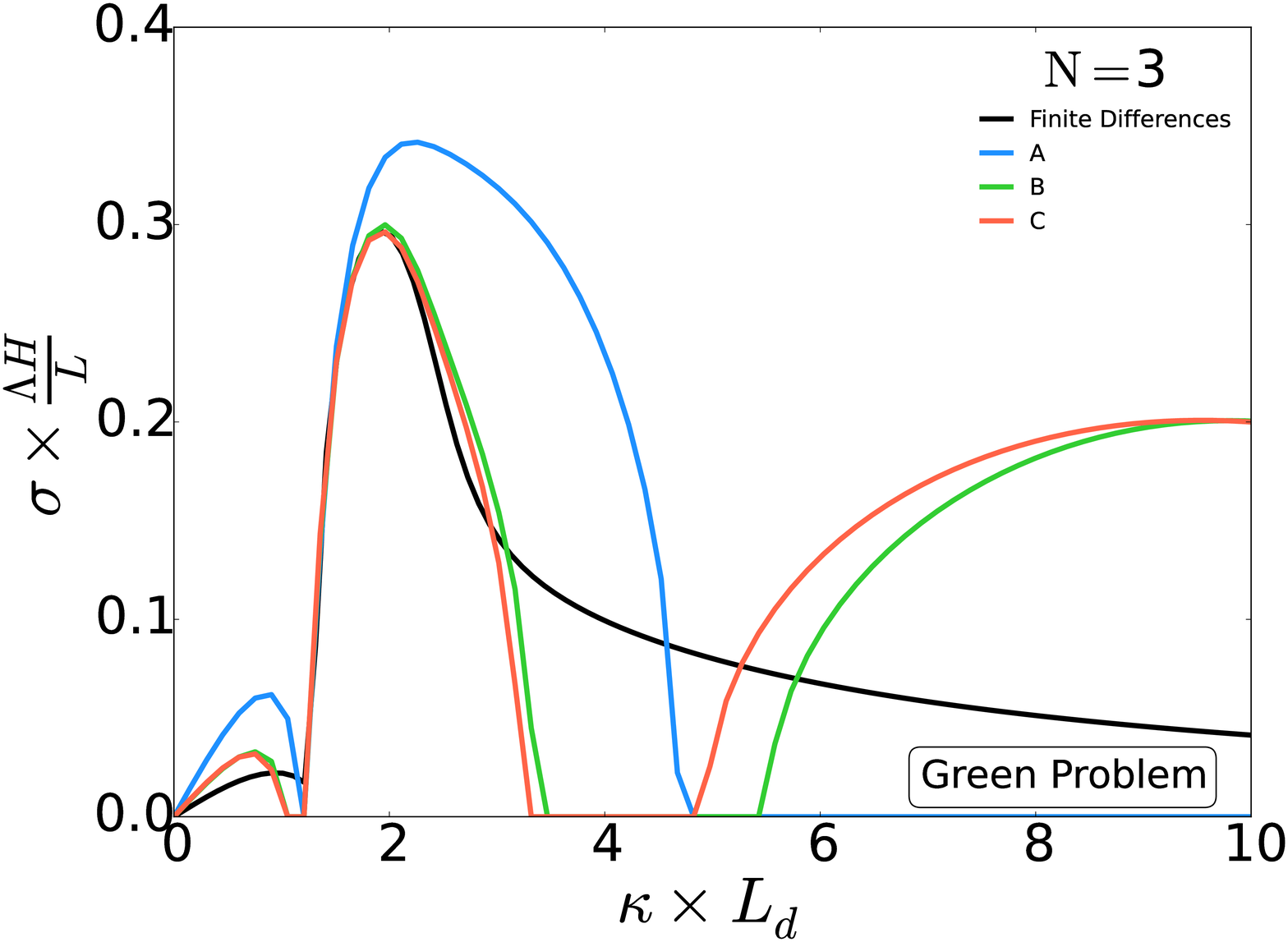} 
\includegraphics[width=12pc,angle=0]{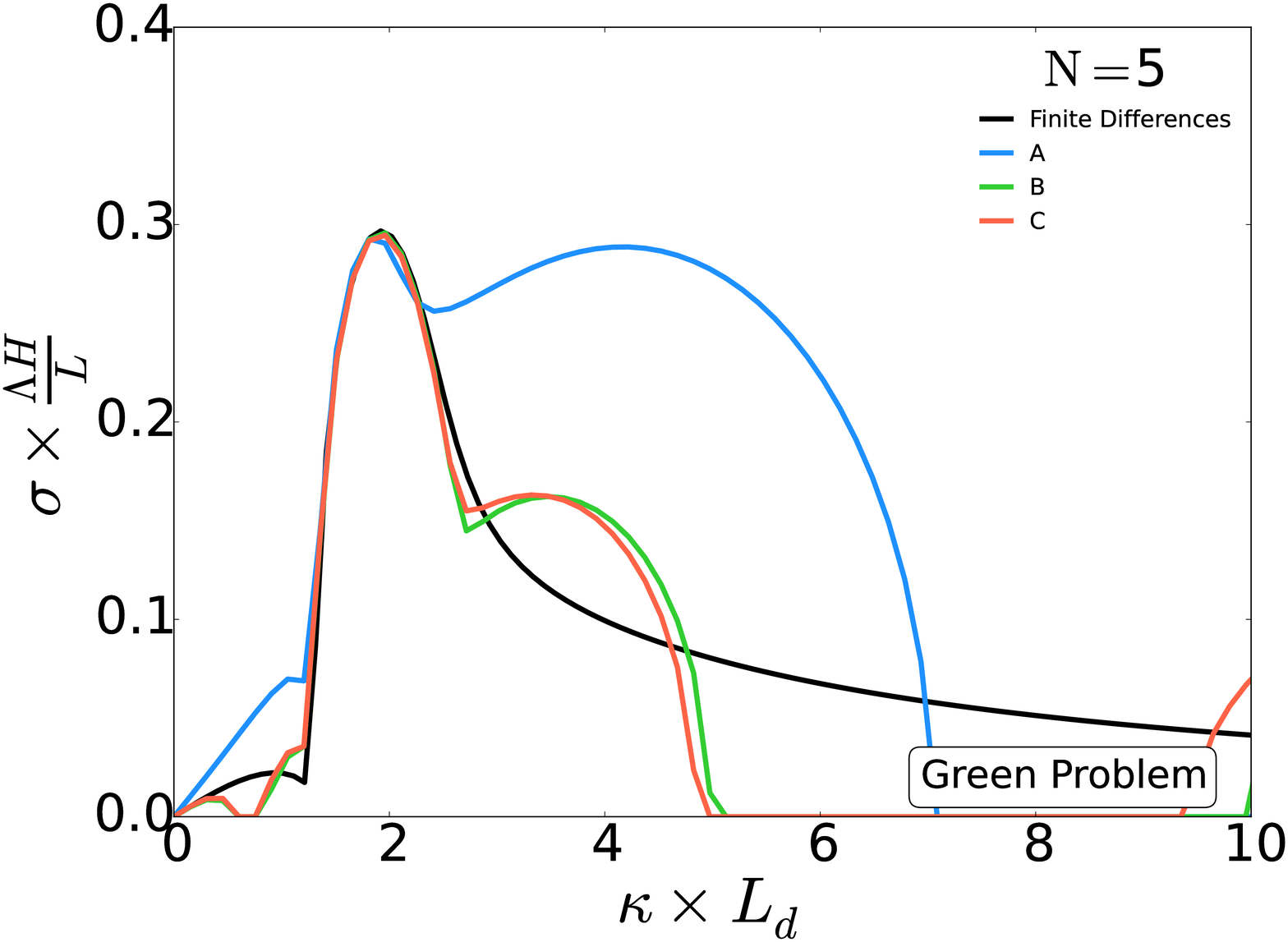}\includegraphics[width=12pc,angle=0]{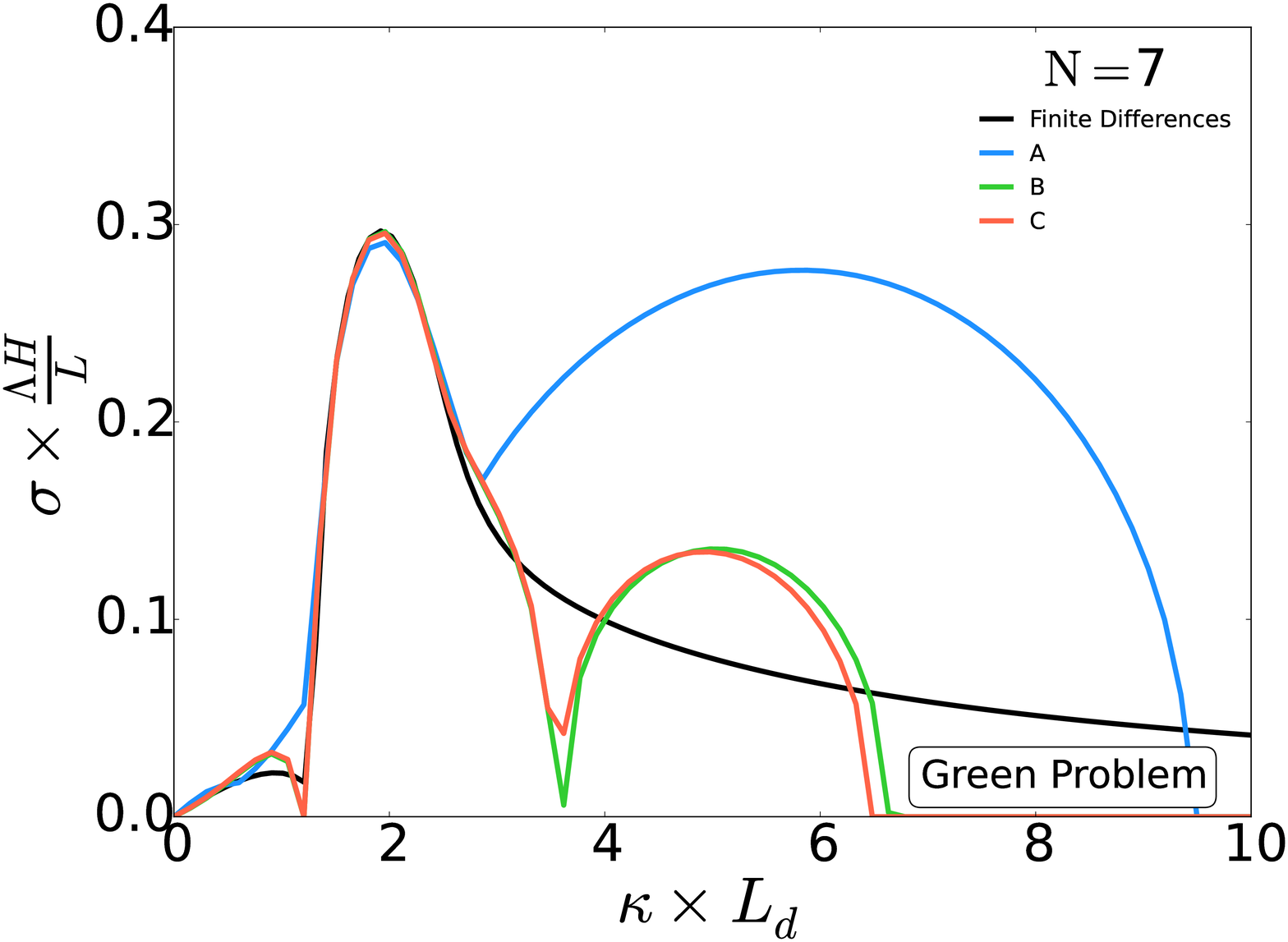} \\
\includegraphics[width=12pc,angle=0]{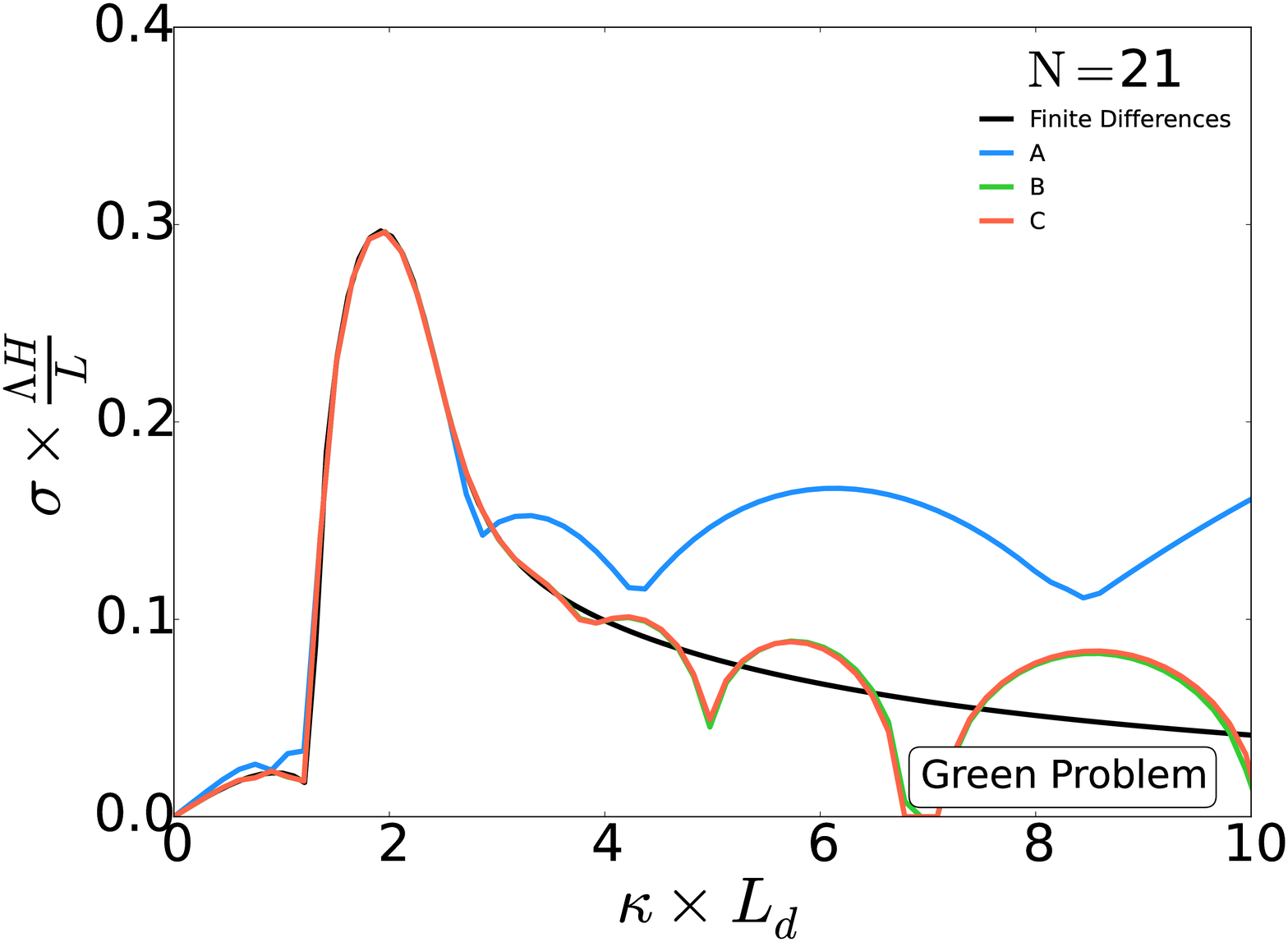}\includegraphics[width=12pc,angle=0]{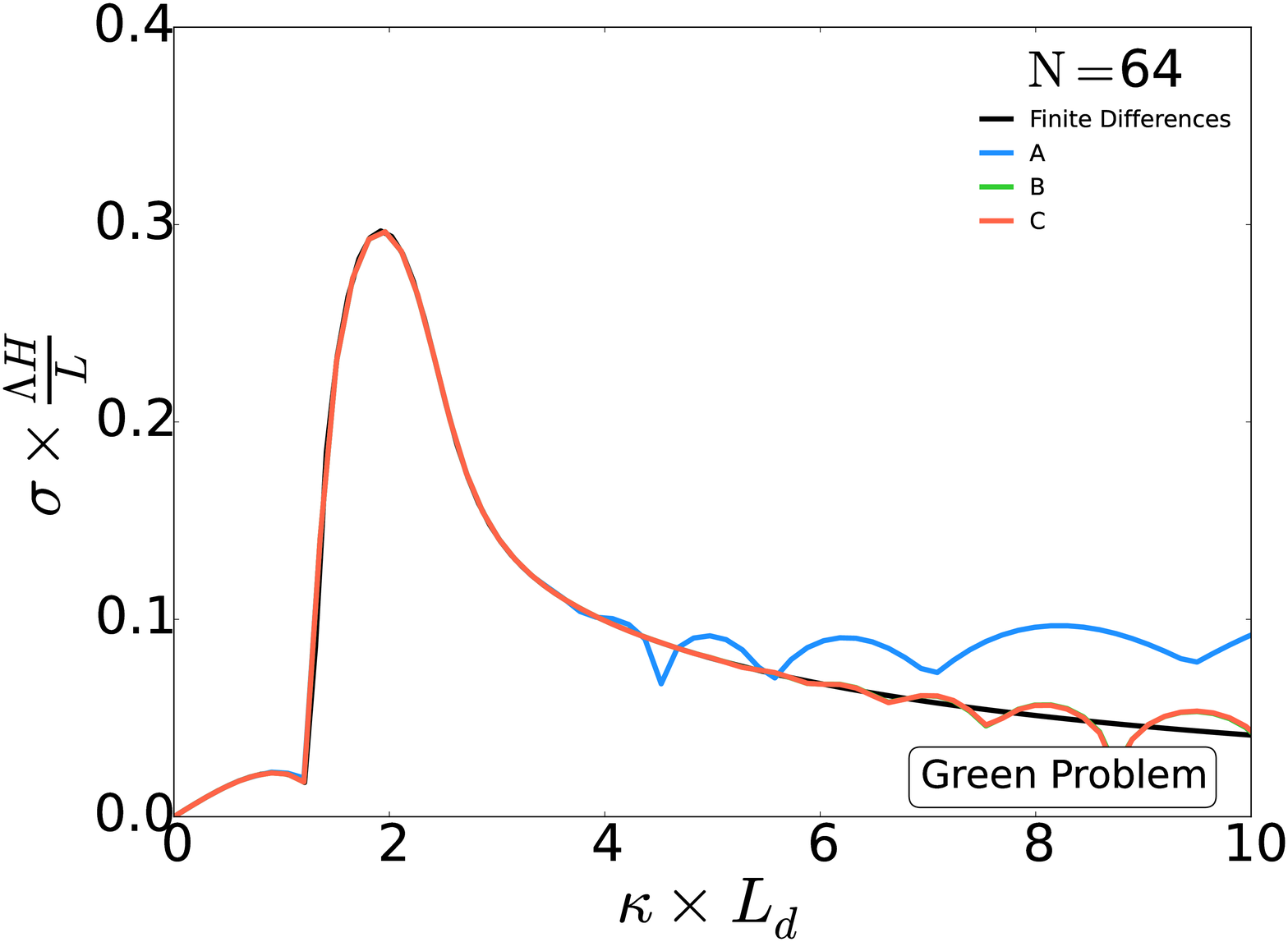}\includegraphics[width=12pc,angle=0]{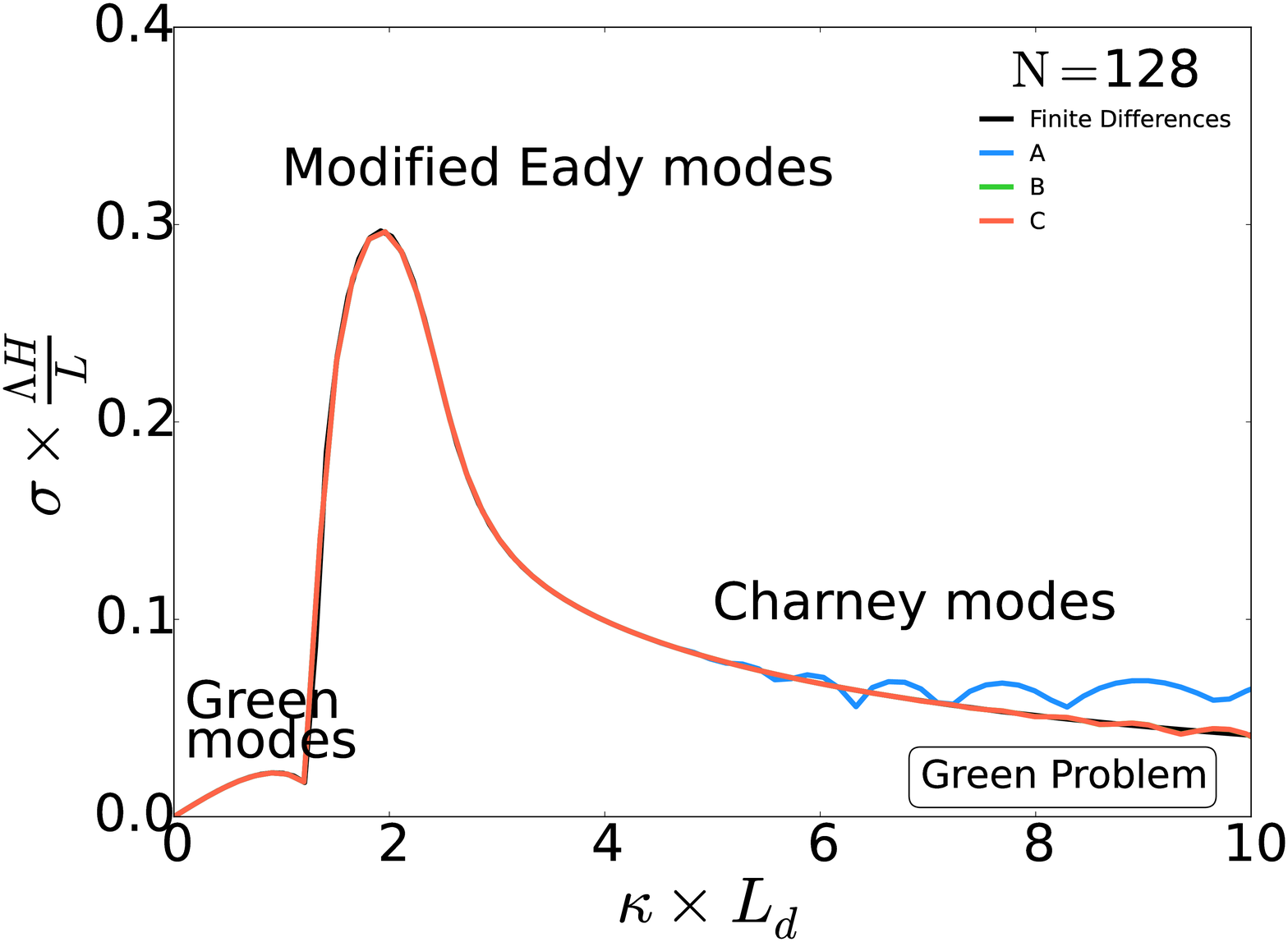}
\end{center}
\caption{Growth rate for the Green problem with $\hat{\beta} = 1$ as a function of the zonal wavenumber ($l=0$) using approximations A, B, C with various number of baroclinic modes ($\nmax$). The black line is a finite-differences solution with 1000 vertical levels.}
\label{beta-eady_growth_rate}
\end{figure*}

To further explore the relative merit and efficiency of approximations A, B, and C we study the instability properties of a system with nonzero $\beta$. For simplicity, we consider a problem with Eady's base-state $\psi = - (1+z)y$ on a $\beta$-plane. This  is similar to the problem originally considered by \cite{green1960} and  \cite{charney1971}. The major difference is that Charney considered a vertically semi-infinite domain \citep{charney1947,pedlosky1987} while we follow Green and consider a finite-depth domain with $-1<z<0$.

We obtain the exact system for this ``Green problem" by linearizing the QG equations \eqref{QGPV1}-\eqref{bc1} about the base-state \eqref{eadyStream} with background PV $\beta y$, where $\hat{\beta}$ is the nondimensional planetary PV gradient. Assuming $\psi = \hat{\phi} \exp[\ii(k \,x + l \,y - \omega \, t)]$, we obtain
\beq
\label{charney_eq}
(U - c)\left[ \hat{\phi}_{zz} -\kappa^2 \hat{\phi}\right] + \hat{\beta} \hat{\phi} = 0\com \qquad -1<z<0 \com
\eeq
and
\beq
\label{charney_bc}
(U-c)\,\hat{\phi}_z - \hat{\phi} = 0\com \qquad z=-1\com0\per
\eeq
 As a reference solution, we solve the eigenproblem \eqref{charney_eq}-\eqref{charney_bc} using a centered second-order finite-difference scheme with 1000 vertical levels: see Figure \ref{beta-eady_growth_rate}.

The Green problem supports three classes of unstable modes, indicated in the lower right   panel ($\nmax = 128$) of Figure \ref{beta-eady_growth_rate}: $(1)$ the ``modified Eady modes'', which are instabilities that arise from the interaction of  Eady-like edge waves,  only slightly modified by $\beta$; $(2)$ the ``Green modes'', which are very long slowly growing modes \citep{vallis2006}; (3) the high-wavenumber ``Charney modes'' are critical layer instabilities that arise from the interaction of the surface edge wave  with the  interior Rossby wave that is supported by nonzero $\beta$.

\subsection{Implementation of approximation A}
The base-state for the Green problem is the same as in the Eady problem. In approximation A, the $\beta$-term adds only a diagonal term to the Eady system  \eqref{eigEadyA} (see appendix C).

\subsection{Implementation of  approximation B}
The base-state is the same as in Eady problem. The steady streamfunction and buoyancy fields that satisfy \eqref{modeEqB} and \eqref{BC_B} exactly are
\beq
\label{base_beta_eady_B}
\Sigma = - \left(1+z\right)y\qquad \text{and} \qquad \Theta^{\pm}= -y\per
\eeq
Assuming $\gq_n = \hat{q}_n(z) \exp[\ii (k \, x + l \, y - \omega^B \,t )]$, the $\nmax+1$ interior equations \eqref{modeEqB} linearized about \eqref{base_beta_eady_B} become
\beq
\label{bets_eady_B_linear}
\sum_{s=0}^{\nmax} \xi_{ns} \hat{q}_s + \hat{\beta} \left(\hat{\phi}_n + \hat{\sigma}_n\right) = c^B_c \, \hat{q}_n\com
\eeq
where
\beq
\label{xi_ns}
\xi_{ns} \defn \tfrac{1}{h} \int_{z^{-}}^{z^{+}}\!\!\!\!\! \sp_n \, \sp_s \, \left(z + 1\right) \dd z\per
\eeq
The  boundary conditions \eqref{BC_B}, linearized about  \eqref{base_beta_eady_B}, become
\beq
\label{bets_eady_B_linear_BC_top}
\hat{\vth}^{+} - \sum_{s=0}^{\nmax} \sp_s^{+}\hat{\phi}_s -\hat{\sigma}^{+} = c^B_c\, \hat{\vth}^{+}\com
\eeq
and
\beq
\label{bets_eady_B_linear_BC_bot}
 - \sum_{s=0}^{\nmax} \sp_s^{-}\hat{\phi}_s - \hat{\sigma}^{-} = c^B_c\, \hat{\vth}^{-}\com
\eeq
where $\hat{\sigma}$ is given by \eqref{sigma_eady}. We use the inversion relationship \eqref{qbreve_B} and the Neumann-to-Dirichlet map \eqref{transfer_func} to recast this eigenproblem into standard form $\sB\,\tilde\sq = c^B \, \tilde\sq$, where $\tilde\sq=[\hat{\vtht},\hat{q}_0,\hat{q}_1,\ldots,\hat{q}_{\nmax-1},\hat{q}_{\nmax},\hat{\vthb}]^\sT$ (see appendix C). 

\subsection{Implementation of  approximation C}
Again the base-state is the same as in the Eady problem. But now there are $\nmax+3$ equations: the two boundary equations of Eady's problem \eqref{bc_linear_C} plus $\nmax+1$ interior equations
\beq
\label{eigCharneyC}
\sum_{m=0}^{\nmax}\sum_{s=0}^{\nmax} \Xi_{nms} \gU_m \, \hat q_s + \hat{\beta} \hat{\psi}_n= c^C \hat q_n\com
\eeq
We use the inversion relationship \eqref{sol111} in \eqref{eigCharneyC} to recast this eigenproblem in the form $\sC \, \tilde\sq = c^C \tilde\sq$, where $\tilde \sq$ is defined as in approximation B (see appendix C).

\subsection{Remarks on convergence}

\begin{figure}[ht]
\begin{center}
\includegraphics[width=19pc,angle=0]{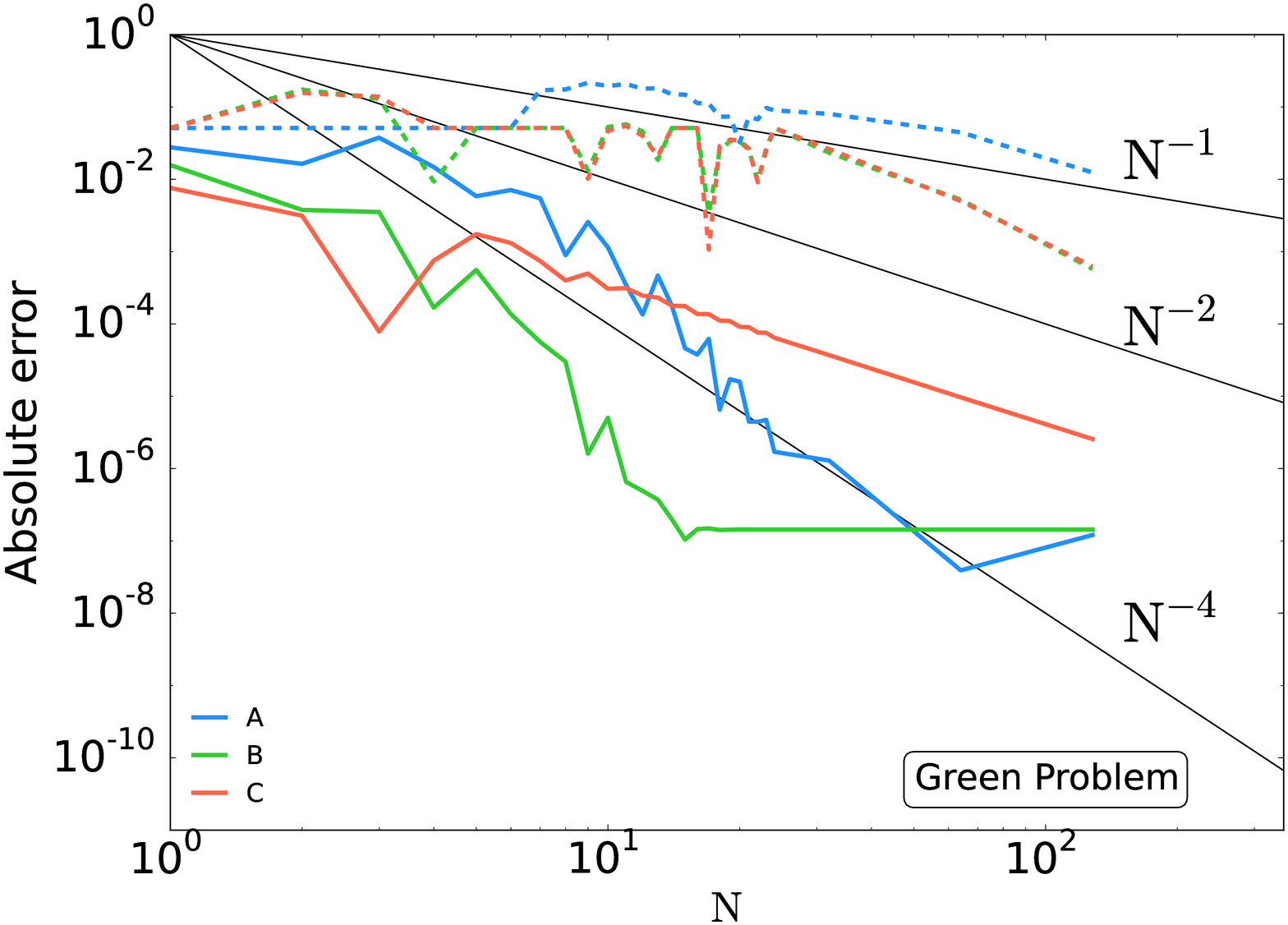}
\end{center}
\caption{Absolute error as a function of number of baroclinic modes ($\nmax$) for the growth rates of the Green  problem. The solid line represent the error at the exact fastest growing mode ($\kappa \approx 1.9$). The dashed line is the error at $\kappa = 8$.}
\label{error_analysis2}
\end{figure}

The most crude truncation ($\nmax = 0$) is stable for approximations A and C. In contrast, the $\nmax = 0$ truncation in approximation B is qualitatively consistent with the modified Eady instabilities: see figure \ref{beta-eady_growth_rate}. With a moderate number of baroclinic modes ($\nmax=2$ or $3$), approximations A, B and C all resolve the modified Eady modes relatively well. At the most unstable modified Eady mode ($\kappa\approx 1.9$), approximation B has typically the smallest error because it solves the surface problem exactly. As in the Eady problem, approximation A converges ($\sim\nmax^{-4}$) faster than approximations B and C ($\sim\nmax^{-2}$) at the most unstable mode, but B and C converge faster at high wavenumbers (\ref{error_analysis2}).

\begin{figure}[ht]
\begin{center}
\includegraphics[width=19pc,angle=0]{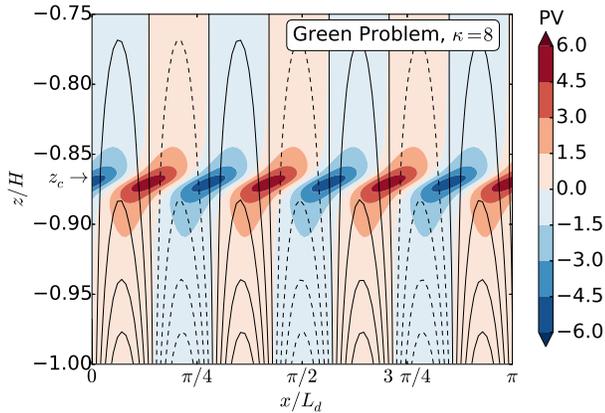}
\end{center}
\caption{Wave structure of the $\kappa = 8$ unstable mode for the Green problem with $\hat \beta = 1$ solved using a second-order finite-difference scheme with 1000 vertical levels. Streamfunction (black contours) and potential vorticity (colors). The streamfunction slightly tilts westward as $z$ increases. The potential vorticity is confined to a small region, the critical layer. The critical level, $z_c$, is the depth where the unstable wave speed matches the velocity of the base-state. Only the bottom quarter of the domain is shown.}
\label{beta-eady_wave_structure}
\end{figure}

Approximations A, B, and C all converge very slowly to the high-wavenumber Charney modes  (Figures \ref{beta-eady_growth_rate} and \ref{error_analysis2}). These modes are interior critical-layer instabilities \citep{pedlosky1987} and the critical layer is confined to a small region about the steering  level (i.e., the depth at which the phase speed matches the base velocity --- see figure \ref{beta-eady_wave_structure}). With finite base-state shear, the critical layer is always in the interior. Thus, the problem is not that standard vertical modes are inefficient because they do not satisfy inhomogeneous boundary conditions; a low resolution finite-difference solution also presents such ``bubbles" in high-wavenumber growth rates  (not shown). Resolution of the interior critical layer, not the surface boundary condition, is a problem for all methods at high wavenumbers. The ``surface-aware'' modes of \cite{smith_vanneste2013} have the same limitation --- a large number of vertical modes is required to resolve the critical-layer instabilities (K. S. Smith, pers. comm.).

For example, with $\nmax<25$, at $\kappa = 8$, approximations are qualitatively inconsistent with the high-resolution  finite-difference solution. For larger values of $\nmax$, the growth rate convergence for approximations B and C scales $\sim\!\nmax^{-3}$. The growth rate for approximation A converges painfully slowly ($\sim\!\nmax^{-1}$). As in the Eady problem, at large wavenumbers, the growth rate for approximation A is qualitatively different from that of  the finite-difference solution because of spurious instabilities associated with the rapidly oscillatory base-state PV gradient.

\section{Summary and conclusions}

The Galerkin approximations A, B, and C  are equivalent if there is are no buoyancy variations at the surfaces. Thus all three approximations are well-suited for applications with zero surface buoyancy \citep{flierl1978,fu_flierl1980,hua_hadivogel1986}. But with nonzero surface buoyancy the three approximations are fundamentally different. 

Approximation A, originally introduced by \cite{flierl1978}, represents the streamfunction as a Galerkin series in standard vertical modes and defines the potential vorticity so that the inversion problem is satisfied  exactly.  The most important limitation  of A is that the interior PV evolves independently of the surface buoyancy (as if $\vthtb$ were zero). The evolution equations in approximation A \eqref{modeFl1} are relatively simple, and the system conserves energy \eqref{cons_E_A} and potential enstrophy \eqref{cons_Ens_A}. But, in approximation A, nonzero surface buoyancy results in  an interior PV that distributionally converges to $\delta$-distributions at the boundaries as $\nmax \to \infty$. These smeared-out Brethertonian $\delta$-functions  provide a very inaccurate representation of the true inhomogeneous surface boundary condition. Finite difference schemes  have a similar pathology \citep{smith2007,tulloch_smith2009a}. As a result of this artificial interior PV gradient,  solutions with a small number of modes are qualitatively misleading and convergence at large wavenumbers is very slow ($\sim\!\nmax^{-1}$). Slow convergence was previously noted by \cite{hua_hadivogel1986} --- see their figure 2. Furthermore, even if heroic values of $\nmax$ achieve convergence at say $\kappa=10$, we conjecture that there will always be spurious unstable modes at even  larger wavenumbers. In some simulations these unphysical high-wavenumber instabilities  might be eliminated by hyperviscosity or by a scale-selective filter. But one must be aware of potential effects on the evolution of the system.

Approximation B, originally introduced by \cite{tulloch_smith2009b} using one baroclinic mode and constant buoyancy frequency, takes the opposite starting point from approximation A. B represents the PV as a Galerkin series in standard modes and calculates the streamfunction that satisfies the exact inversion problem associated with the approximate PV. The linear inversion problem can be split into an interior contribution with homogeneous boundary conditions and a surface contribution with zero interior source and inhomogeneous boundary conditions \citep{lapeyre_klein2006,tulloch_smith2009b}. Thus the exact interior streamfunction associated with the approximate PV is a Galerkin series, but the surface streamfunction must be computed using other methods.  Because the surface streamfunction projects onto the interior solution the energy is not diagonalized. Indeed, approximation B conserves neither  energy nor potential enstrophy if  $\beta\neq 0$.

Approximation C represents \textit{both} the PV and the streamfunction by Galerkin series. Although the inversion problem is not satisfied exactly, the relation between modal streamfunction $\gpsi_n$ and PV $\gq_n$ is obtained by Galerkin projection of the exact inversion relationship and prominently exhibits  the surface buoyancy  fields: see \eqref{sol111}.  Approximation C is the most consistent  because it uses the same level of approximation for both PV and streamfunction. The evolution equations \eqref{modeEqC}   are relatively simple, and the approximate system conserves total energy \eqref{cons_E_C}. The most vexing limitation of C is the lack of potential enstrophy conservation with $\beta \neq 0$. But this does not mean that approximation A is better than approximation C: all approximations conserve potential enstrophy when $\beta=0$, or if $\vthtb = 0$, and none conserve an  analog of the exact potential enstrophy \eqref{cons_Ens_A} when $\beta\neq0$ and $\vthtb\neq0$. 

With nonzero interior PV gradients  the convergence of all  approximations is  slow for the high-wavenumber Charney-type modes. The critical layer associated with these modes spans a very small fraction of the total depth (Figure \ref{beta-eady_wave_structure}). To accurately resolve these near-singularities at the steering level there is no better solution than having high vertical resolution in the interior. 

For problems with nonuniform surface buoyancy and nonzero interior PV gradient, we recommend approximation C for obtaining  solutions to the three-dimensional QG equations using standard vertical modes.

The codes that produced the numerical results of this paper, plotting scripts, and supplementary figures are openly available at \texttt{github.com/crocha700/qg$\_$vertical$\_$modes}.

%
\acknowledgments
CBR is grateful for a helpful conversation with G. R. Flierl. We thank Joe LaCasce and Shafer Smith for reviewing this paper. Geoff Vallis pointed out that Green (1960) first considered the Eady$+\beta$ problem. This research was supported by the National Science Foundation under award  OCE 1357047.

%
\appendix[A]
\appendixtitle{On the convergence of Galerkin series in standard modes}

\cite{jackson1914} gives conditions for the uniform convergence of series expansions in eigenfunctions of the Sturm-Liouville eigenproblem
\beq
\label{jack1}
\frac{\dd^2 \sP_n}{\dd Z^2} + \left[\rho^2_n - \Lambda(Z)\right] \sP_n = 0 \com
\eeq
defined on the interval $Z\in[0,\pi]$ with boundary conditions
\beq
\label{jack2}
\sP_n'(0) - \gamma_0 \sP_n(0) = 0\com \;\; \text{and} \;\; \sP'_n(\pi) - \gamma_\pi \sP_n(\pi) = 0
\eeq
where $\gamma_0$ and $\gamma_\pi$ are real constants of arbitrary sign and $\rho_n^2$ is the eigenvalue.
The equations defining the standard modes \eqref{SL1}-\eqref{SL2} can be brought to this form using the following Liouville transformation
\beq
Z(z) = \frac{1}{\bar{Z}}\int_{\zm}^{z} S(\xi)^{-1/2} \dd \xi\com \;\; \text{with} \;\; \bar{Z} 
\defn \frac{1}{\pi}\int_{\zm}^{\zp}S(\xi)^{-1/2} \dd \xi\com
\eeq
and 
\beq
\sP_n(Z)= S(z)^{1/4}\sp_n(z)\com \quad \text{where} \quad S(z) \defn \frac{f_0^2}{N^2(z)}\per
\eeq
The eigenvalues are related by $\rho_n = \bar{Z} \kappa_n$ and
\beq
\Lambda(Z) = \bar{Z}^2 \left[\frac{1}{4}\frac{\dd^2 S}{\dd z^2}-\frac{1}{16S}\left(\frac{\dd S}{\dd z}\right)^2\right]\per
\eeq
The boundary condition for the standard modes \eqref{SL2} implies that the transformed modes satisfy \eqref{jack2} with
\beq\gamma_0 = \frac{4S(\zm)^{1/2}}{\bar{Z}\dd S(\zm)/\dd z}\com\quad \text{and}\quad \gamma_{\pi} = - \frac{4S(\zp)^{1/2}}{\bar{Z}\dd S(\zp)/\dd z}\per
\eeq
If $\dd S/\dd z=0$ at a boundary then the appropriate condition at that boundary is $P_n=0$.

A special case of Theorem I from \cite{jackson1914} states that the expansion of a function $\phi(Z)$ as a series in eigenfunctions $\sP_n$ converges absolutely and uniformly provided that both $\dd\phi/\dd Z$ and $\dd\Lambda/\dd Z$ are continuous and bounded, regardless of whether or not $\phi$ satisfies the same boundary conditions as $\sP_n$.
(The remainder of the theorem concerns the rate of convergence under stronger conditions on $\phi$ and $\Lambda$.)
The streamfunction, potential vorticity, and buoyancy frequency profiles are typically assumed to be smooth in studies of QG dynamics, which implies that both $\phi$ and $\lambda$ will satisfy the above conditions.
Uniform convergence over $Z\in[0,\pi]$ implies uniform convergence over $z\in[\zm,\zp]$.

\appendix[B]
\appendixtitle{Details of the derivation of quadratic conservation laws for approximate equations \label{appA}}
\subsection{Approximation A}
To obtain the energy conservation in approximation A we multiply the modal equations \eqref{modeFl1} by $-\gpsi_n$, integrate over the horizontal surface, and sum of on n, to obtain
\begin{align}
\label{E_cons_A_1}
&\frac{\dd}{\dd t} \sum_{n=0}^{\nmax}\int \half \left[ (\nabla\breve{\psi_n})^2 + \kappa_n^2\breve{\psi}_n^2 \right] \dd S \nonumber \, + \\ & \sum_{n=0}^{\nmax} \sum_{m=0}^{\nmax} \sum_{s=0}^{\nmax}  \Xi_{nms} \int \,\gpsi_n\,  \sJ\left(\gpsi_m,\helms\gpsi_s\right) \dd S = 0\per
\end{align}
Notice that
\begin{align}
    \int \gpsi_n \,\sJ\left(\gpsi_m,\helms\gpsi_s\right)\, \dd S = \int \helms\gpsi_s\,  \sJ\left(\gpsi_n,\gpsi_m\right) \,\dd S\per
\end{align}
Hence the triple sum term in \ref{E_cons_A_1} vanishes identically because $\Xi_{nms}$ is fully symmetric and the Jacobian is skew-symmetric. Thus we obtain conservation of energy \eqref{cons_E_A}.  Similarly, to obtain the potential enstrophy conservation  law in \eqref{cons_Ens_A} we multiply the modal equations \eqref{modeFl1} by $\helmn\gpsi_n$, integrate over the surface, sum on $n$, and invoke the same symmetry arguments used for the energy conservation.

\subsection{Approximation B}

\subsubsection*{Energy nonconservation}
The analog of \eqref{E_defn} in approximation B is 
\beq
\label{Edefn_B}
E^B_N \defn E_{\phi} + E_{\sigma} + E_{\phi\sigma}\per
\eeq
The three terms in \eqref{Edefn_B} are
\begin{align}
\label{Ephi_defn}
E_{\phi} &=  \tfrac{1}{h} \int  \left[ \left| \nabla \phiB \right|^2 + \bur \left(\p_z \phiB\right)^2 \right] \dd V \nonumber \\
               &  =   \sum_{n=0}^{\nmax}\,  \int \halfrho  \left[ \left| \nabla \gphi_n \right|^2 + \kappa_n^2\gphi_n^2 \right] \, \dd S \com
\end{align}
\beq
\label{Esig_defn}
E_{\sigma} = \tfrac{1}{h} \int  \halfrho  \left[ \left| \nabla \sigma \right|^2 + \bur \left( \p_z \sigma\right)^2 \right] \dd V,
\eeq
and
\begin{align}
    \label{Ephisig_defn}
    E_{\phi\sigma} &=  \tfrac{1}{h} \int \rz  \left[ \nabla \phiB \bcdot \nabla \sigma  + \bur \p_z \phiB\, \p_z \sigma \right] \dd V,\nonumber \\
                     & =  \sum_{n=0}^{\nmax} \int  \rz  \gsigma_n \helmn\gphi_n \, \dd S \per
\end{align}
The cross-term  $E_{\phi\sigma}$ is not zero because the surface streamfunction $\sigma$ projects on the standard vertical modes i.e., $\gsigma_n$ is nonzero. To obtain an equation for $E_B$ we form evolution equations for the three components in \eqref{Edefn_B} and add them. The final result is 
\begin{align}
    \label{Eeqn_B}
    \frac{\dd E^B_N}{\dd t} &=  \sum_{n=0}^{\nmax} \sum_{s=0}^{\nmax}   \tfrac{1}{h} \int\!\! \sp_n \sp_s  ( \gphi_n+\breve \sigma_n) \sJ(\sigma,\gq_s) \dd V \nonumber \\
                            &  + \sum_{m=0}^{\nmax} \sum_{n=0}^{\nmax} \sum_{s=0}^{\nmax} \Xi_{mns} \int  \  \gsigma_n \sJ(\gphi_m,\helms\gphi_s)\, \dd S \per
\end{align}
The simplest model with barotropic interior dynamics ($\nmax = 0$) conserves energy. With richer interior structure, however, the right-hand-side of \eqref{Eeqn_B} is generally nonzero. Consider the ``two surfaces and two modes'' (TMTS) model of \cite{tulloch_smith2009b}, corresponding to $\nmax =1$ with constant buoyancy frequency. Using nondimensional variables  the energy equation \eqref{Eeqn_B} becomes
 \beq
\label{Eb_tmts}
\frac{\dd E^B_1}{\dd t} = \int\left[ \phi_1 \,\sJ\left(\gsigma_1\com \triangle_1\gphi_1\right) - \tfrac{1}{\sqrt{2}} \triangle_1\gphi_1\,\sJ\left(\gsigma_1\com\gsigma_2\right)\right] \dd S\per
\eeq
We now construct an example in which we can analytically show that the right-hand-side of \eqref{Eb_tmts} is nonzero. This example should be interpret as an initial condition for which energy is guaranteed to grow or decay. For simplicity we consider $\triangle\gphi_1 = \lambda \gphi_1$, where $\lambda$ is a constant, so that the first term on the right-hand-side of \eqref{Eb_tmts} is identically zero. As for the surface streamfunction, we choose
\beq
\label{surface_tmts}
\sigma = \frac{\cosh{(z+1)}}{\sinh 1}\cos x +  \frac{\cosh{z}}{\sinh1}\sin x\per
\eeq
We use $\phi_1 =  \sin x\,\cos y$ so that all fields are periodic with the same period. Integrating over one period in both directions gives 
\beq
\label{energy_noncons_tmts}
\frac{\dd E^B_1}{\dd t} = \frac{\lambda}{\sqrt{2}\,(1 + 5\pi^2 + 4\pi^4)} \neq 0\per
\eeq
The total energy $E^B_1$ grows or decays depending on the sign of $\lambda$. Thus the analog of the exact energy \eqref{E_defn} is not generally conserved in approximation B. 

\subsubsection*{potential enstrophy nonconservation}
The analog of the exact potential enstrophy \eqref{ensdens} is not conserved in approximation B. We attempt to form a potential enstrophy conservation by multiplying the interior equations \eqref{modeEqB} by $\helmn\gphi_n$ and integrating over the surface, and summing on $n$, 
\begin{align}
\label{enstrophy_deriv_B_1}
\frac{\dd}{\dd t} \sum_{n=0}^{\nmax} \int \half (\helmn\gphi_n)^2 \dd S - \beta \sum_{n=0}^{\nmax} \int \helmn \gsigma_n \p_x \gphi_n \dd S  = 0\per
\end{align}
The potential enstrophy given by the sum on the left-hand-side of \eqref{enstrophy_deriv_B_1} is conserved in the special case $\beta=0$.  For $\beta\neq0$ we first form an equation for $\helmn \gsigma_n$, and then cross-multiply with the modal equations \eqref{modeEqB}, integrate over the surface, and sum on $n$. Eliminating the $\beta$-term in \eqref{enstrophy_deriv_B_1} yields
\begin{align}
\label{enstrophy_noncons_B}
\frac{\dd}{\dd t}& \sum_{n=0}^{\nmax} \half (\helmn \gphi_n)^2 + (\helmn \gsigma_n)\gq_n \dd S = \nonumber \\ &- \sum_{m=0}^{\nmax}\sum_{n=0}^{\nmax}\sum_{s=0}^{\nmax} \Xi_{mns} \int \helmn\gsigma_n \sJ\left(\gphi_m\com\helms\gphi_s\right) \dd S \nonumber \\ & - \sum_{n=0}^{\nmax} \tfrac{1}{h} \int \helmn\gsigma_n \sp_n\sp_n \sJ\left(\sigma\com\helms\gphi_s\right) \dd V \nonumber \\ & +
 \sum_{n=0}^{\nmax}\sum_{m=0}^{\nmax} \tfrac{1}{h} \int \helmn \gphi_n\,\Big[ \sp_n^{+}\sJ\left(\sigma^{+} + \sp_m^{+}\gphi_m\com\vtht\right) \nonumber \\ & ~~~~~~~~~~~~~~~~~~~~~~~~~~~~ -\sp_n^{-}\sJ\left(\sigma^{-}+\sp_m^{-}\gphi_m\com\vthb\right)\Big] \dd S\per
\end{align}
The right-hand-side of \eqref{enstrophy_noncons_B} is nonzero even in the simplest model ($\nmax=0$). 

\subsection{Approximation C}
To obtain the conservation of energy in approximation C we multiply the modal equations \eqref{modeEqC} by $-\gpsi_n$, integrate over the horizontal surface, and sum on $n$, to obtain
\begin{align}
\label{E_cons_C_1}
\frac{\dd}{\dd t} &\sum_{n=0}^{\nmax}\int \left[ (\nabla\gpsi_n)^2  +\kappa_n^2\gpsi_n^2 \right] \dd S \nonumber \, \\&-   \sum_{n=0}^{\nmax}\tfrac{1}{h}\int \gpsi_n  \p_t\, (\sp^{+}_n\vtht - \sp^{-}_n\vthb)\, \dd S \,\nonumber \\ &+   \sum_{n=0}^{\nmax} \sum_{m=0}^{\nmax} \sum_{s=0}^{\nmax}  \Xi_{nms} \int \rz \gpsi_n\,  \sJ\left(\gpsi_m, \helms\gpsi_s\right)\, \dd S = 0\com
\end{align}
The triple sum term vanishes by the same symmetry arguments used above in approximation A. The term on the second line of \eqref{E_cons_C_1} is also zero:  multiply the boundary conditions \eqref{BC77} by $\sp_n^{\pm}\,\gpsi_n$ and integrate over the horizontal surface. Thus we obtain the  energy conservation law in \eqref{cons_E_C}.

\subsubsection*{Potential enstrophy nonconservation}
As in approximation B, the analog of the exact potential enstrophy \eqref{ensdens} is not conserved in approximation C. The potential enstrophy equation with $\beta\neq0$ is formed analogously to the approach used above in approximation B. The final result is
\begin{align}
\label{deriv_enstrophy_C_2}
&\frac{\dd}{\dd t}\sum_{n=0}^{\nmax} \int \frac{\gq_n^2}{2}  + \gq_n\helmn \gsigma_n\, \dd S =  \nonumber \\&+ \sum_{m=0}^{\nmax}\sum_{n=0}^{\nmax} \tfrac{1}{h} \int\!\!  \gq_n \sp_n^{+}\sp_m^{+} \sJ\left(\gpsi_m,\vtht\right) - \gq_n \sp_n^{-}\sp_m^{-} \sJ\left(\gpsi_m,\vthb\right)  \nonumber \\ & - \sum_{m=0}^{\nmax}\sum_{n=0}^{\nmax}\sum_{s=0}^{\nmax} \int \Xi_{mns} \helmn\gsigma_n \sJ\left(\gpsi_m\com\gq_s\right)\per
\end{align}
The right-hand-side of \eqref{deriv_enstrophy_C_2} is zero for the simplest model ($\nmax=0$), but it is generally nonzero.

\appendix[C]
\appendixtitle{Details of the stability  problems \label{appB}}

\subsection{The interaction tensor}
Because the standard vertical modes with constant stratification are simple sinusoids \eqref{std_modes_constN}, the interaction  coefficients  \eqref{Xidef} can be computed analytically. First we recall that $\Xi_{ijk}$ is fully symmetric. Permuting the indices so that $i \ge j \ge k$ we obtain
\beq
\label{Xi_ijk}
\vspace{.1cm}
\Xi_{ijk} = \begin{cases} \,\,\,\,1: & i=j,\,k=0\,;\\
\frac{\sqrt{2}}{2}: & i = j+k\,;\\
\vspace{.1cm}
                   \,\,\,\, 0: & \text{otherwise}\per
            \end{cases}
\eeq
The second line in \eqref{Xi_ijk} corrects a factor of $\tfrac{1}{2}$ missed by \cite{hua_hadivogel1986}.

\subsection{Approximation A}

Using the symmetry in $\Xi_{nms}$, and the inversion relation \eqref{invr_A}, we rewrite  row $n+1$ of the linear Green system
\begin{align}
\label{eigBEadyA2}
\sum_{s=0}^{\nmax} \sum_{m=0}^{\nmax} \Xi_{nms} \left(\gU_m + \p_y \gQ_m\,\al_s\right)   \hat q_s + \hat{\beta} \al_n  \hat{q}_n = c^A \hat q_n, 
\end{align}
where the inverse of the $n$'th mode Helmholtz operator in Fourier space is 
\beq
\label{bare_rws}
\al_n \defn -(\kappa^2 + (n\pi)^2)^{-1}\per
\eeq
The Eady problem is the special case $\hat{\beta} = 0$. We use a standard eigenvalue-eigenvector algorithm to obtain the approximate eigenspeed $c^A$.

\subsection{Approximation B}
The Green eigenvalue problem in  \eqref{bets_eady_B_linear} through \eqref{bets_eady_B_linear_BC_bot} can be recast in the standard form $\sB\,\sq = c^B \, \sq$, where  $\tilde\sq=[\hat{\vtht},\hat{q}_0,\hat{q}_1,\ldots,\hat{q}_{\nmax-1},\hat{q}_{\nmax},\hat{\vthb}]^\sT$. The first and last rows of the system stem from the boundary conditions \eqref{bets_eady_B_linear_BC_top}-\eqref{bets_eady_B_linear_BC_bot}
\beq
\label{first_row_Bb}
\left(1 - \frac{\coth{\kappa}}{\kappa}\right)\! \hat{\vth}^{+} - \sum_{s=0}^{\nmax}\! \sp_s^{+} \al_s\hat{q}_s - \frac{\cosech{\kappa}}{\kappa} \,\hat{\vth}^{-} = c^B\, \hat{\vth}^{+}\com
\eeq
and
\beq
\label{last_row_Bb}
\frac{\cosech{\kappa}}{\kappa}\, \hat{\vth}^{+}- \sum_{s=0}^{\nmax} \!\sp_s^{-} \al_s \, \hat{q}_s + \frac{\coth{\kappa}}{\kappa}\,\hat{\vth}^{-}  = c^B\, \hat{\vth}^{-}\per
\eeq

The $(n+1)$'th row originates from the $n$'th interior equation \eqref{bets_eady_B_linear}
\begin{align}
    \label{n_p_one}
    - \hat{\beta}\,\sp_n^{+}\, \al_n\, \vtht +  \sum_{s=0}^{\nmax} \gamma_{ns} \hat{q}_s + \left(\beta \al_n + 1 \right)  + \hat{\beta}\,\sp_n^{-}\, \al_n\, \vthb = c^B \, \hat{q}_n\com
\end{align}
where the symmetric matrix $\gamma_{ms}$ is
\beq
\label{gamma_ns}
\gamma_{ij}  \defn  \int_{-1}^{0} \sp_i \, \sp_j \, z  \, \dd z = \begin{cases}
\vspace{.1cm}    
    \,\,\,\,\,\,\,\,\,\,  -\tfrac{1}{2}:  & i = j\,;\\
\vspace{.1cm}    
      \,\,\, \,\,\,\,\, \,\,\frac{2\sqrt{2}}{(j \, \pi)^2}: & i = 0,\, j\,\, \text{ is odd}\,;\\
\tfrac{4\,\left(i^2 + j^2\right)}{\left[\left(i^2 - j^2\right) \, \pi\right]^2}: &\,\,i+j\,\text{ is odd}\per
\end{cases}
\eeq

\subsection{Approximation C}

\subsubsection*{The Eady problem}
The $2\times2$ eigenproblem is
\beq
\label{modal_matrix}
\underbrace{
\begin{bmatrix*} 
    {\UGN}^{+} + \Sigma_{\nmax} & -\Omega_{\nmax}\\
    \Omega_{\nmax} & {\UGN}^{-}-\Sigma_{\nmax}
\end{bmatrix*}}_{\defn \sC}\begin{bmatrix*} 
\vththat\\
\vthbhat
\end{bmatrix*} = c^C \begin{bmatrix*} 
\vththat\\
\vthbhat
\end{bmatrix*}\com
\eeq
where
\beq
\label{Ss_defn}
\Sigma_{\nmax} \defn \al_0+ 2\sum_{n=1}^{\nmax}  \al_n\com 
\quad \text{and} \quad
\Omega_{\nmax} \defn \al_0 + 2\sum_{n=1}^{\nmax}(-1)^n  \al_n\per
\eeq
For finite $\nmax$, the approximate eigenspeed is
\begin{align}
\label{eigs_C}
&c^C=  \frac{{\UGN}^+ + {\UGN}^-}{2} \pm \Bigg[\bigg( \frac{{\UGN}^+ + {\UGN}^-}{2} \bigg)^2 \nonumber \\ & -{\UGN}^+{\UGN}^- + ({\UGN}^+-{\UGN}^-)\Sigma_{\nmax} + \Sigma_\nmax^2 - \Omega_\nmax^2 \Bigg]^{1/2}\per
\end{align}
The sums \eqref{Ss_defn}  become exact in the limit $\nmax\to\infty$  
\beq
\label{coth_series}
\Sigma_{\infty} = -\frac{\coth\kappa}{\kappa}\com \qquad \text{and}\qquad \Omega_{\infty} = -\frac{\cosech\kappa}{\kappa}\per
\eeq
The base velocity also converges to the exact result. Notice that
\beq
1 + \frac{1}{3^2} + \frac{1}{5^2} + \ldots = \sum_{k=1}^\infty \frac{1}{(2k -1)^2} =   \frac{\pi^2}{8}\com
\eeq
and therefore
\beq
{\UG_{\infty}}^{+}= 1\com \qquad \text{and} \qquad {\UG_{\infty}}^{-}= 0\per
\eeq
Thus 
\beq
\label{Mn_Mana}
\sC \to \sB \qquad \text{as} \qquad \nmax \to \infty\com
\eeq
and  the eigenvalues of the Eady problem using approximation C become exact i.e., $c^C \to c^B$ as $\nmax\to\infty$. 

\subsubsection*{The Green problem}
The $(\nmax+3)\times(\nmax+3)$ eigenproblem is
\beq
\label{eig_charney_C}
\sC\,\tilde\sq = c^C \,\tilde\sq\com
\eeq
where $\tilde\sq$ is defined as above in approximation B. The first and last rows of \eqref{eig_charney_C} stem from the boundary conditions \eqref{bc_linear_C}
\beq
\label{first_row_C_beta}
\left({\UGN}^+ + \Sigma_{\nmax}\right) \vththat - \sum_{n=0}^{\nmax} \al_n \sp_n^+ \hat{q}_n - \Omega_\nmax \vthbhat = c^C \vththat \com 
\eeq
and
\beq
\label{last_row_C_beta}
\Omega_\nmax \vththat - \sum_{n=0}^{\nmax} \al_n \sp_n^- \hat{q}_n + \left({\UGN}^- - \Sigma_\nmax\right) \vthbhat = c^C \vthbhat \per
\eeq
Row $n+1$  originates from the $n$'th modal equation \eqref{eigCharneyC}:
\begin{align}
\label{general_row_C_beta}
\hat{\beta} \al_n \sp_n^+ \, \vththat + \sum_{s=0}^{\nmax}\sum_{m=0}^{\nmax}\Xi_{mns}\gU_m\hat{q}_s + \hat{\beta}\, \al_n\,\hat{q}_n  \nonumber \\ - \hat{\beta}\al_n \sp^{-} \, \vthbhat = c^C \hat{q}_n  \per
\end{align}

\bibliographystyle{ametsoc2014}
\bibliography{refs_qg_galerkin.bib}

\end{document}